\def\R{{\rm I}\!{\rm R}}
\def\Z{{\rm Z}\!\!{\rm Z}}

\def\carre{\hbox{\vrule \vbox to 7pt{\hrule width 6pt \vfill \hrule}\vrule
}}

\def\og{\leavevmode\raise.3ex\hbox{$\scriptscriptstyle\langle\!\langle$}}

\def\fg{\leavevmode\raise.3ex\hbox{$\scriptscriptstyle\,\rangle\!\rangle$}}

\font\un = cmbx10 at 14pt

 \null

 \vskip 1cm
 \centerline {\un
Dynamics and Lieb-Robinson estimates }
\medskip
\centerline {\un for lattices of interacting  }
\medskip
\centerline {\un anharmonic oscillators.}
\bigskip
\centerline {\it In the memory of Andrzej Hulanicki.}
\bigskip
 \centerline { L. Amour\footnote
 {$^{(1)}$}
 {laurent.amour@univ-reims.fr}, P. L\'evy-Bruhl\footnote
 {$^{(2)}$}{pierre.levy-bruhl@univ-reims.fr}, J. Nourrigat\footnote
 {$^{(3)}$}{jean.nourrigat@univ-reims.fr} }
\medskip
\centerline {D\'epartement de Math\'ematiques, FRE 3111}
\medskip
\centerline {Universit\'e de Reims. B.P. 1039. 51687 Reims Cedex
2. France}

 \vskip 1cm

 \noindent {\bf Abstract.}  For a class of infinite lattices of
 interacting anharmonic oscillators, we study the existence of the
 dynamics, together with Lieb-Robinson bounds, in a suitable algebra
 of observables.

\bigskip
\noindent {\bf 1. Introduction. Statement of results.}

\bigskip

Infinite lattices of nearest-neighbors interacting harmonic oscillators
are a usual model in quantum statistical mechanics. Among the objects
associated to this model, an important one is the dynamics describing the time evolution
of some algebra of observables, related to the lattice. Such  dynamics on a
lattice was defined by Malyshev-Minlos [MA-MI] and by Thirring [TH],
when the potential is a quadratic form.
\medskip

We also note that, for bounded  Hamiltonian  models, Lieb and
Robinson have established in [LI-R] an estimate, concerning the
propagation speed for the correlation between two local
observables. For these models, the existence of the dynamics is
proven, for example in [NOS], in some algebra (not the same as in
[MA-MI] or (TH]).
\medskip

More recently,
Nachtergaele, Raz, Schlein and  Sims [NRSS] have derived   Lieb-Robinson
type inequalities for lattices of harmonic oscillators with quadratic
interactions with, moreover,  on each site of the lattice, a
self-interaction potential in a more general class.
More precisely, Lieb-Robinson type inequalities are proved ([NRSS])
for Hamiltonians associated to a finite subset $\Lambda$ of the lattice,
and hold uniformly in $|\Lambda|$.  However, to the best of our knowledge,
the existence of  dynamics as
$|\Lambda|\rightarrow \infty$  is  established  when the potential
is a quadratic form, but not  with smaller perturbations.
\bigskip

The aim of this article is twofold. On the first side, we define
a $C^{\star}-$algebra ${\cal W}_2$ which seems to be more convenient, when
the perturbation is turned on, than the Weyl algebra defined in
 [MA-MI] or in [TH], or than the one used in [NOS]. In particular, we
prove the existence of a dynamics (defined as a limit when the number of
sites goes to infinity) for local and non local observables in this
algebra.  On the other side, we are able to perturb the quadratic
potential of interaction in a more general way than in [NRSS], with
not only self-interacting terms. In this
framework,  we also obtain the Lieb-Robinson type inequalities, with a
bound for the propagation speed of the correlations.

\bigskip
We make the choice here to consider a one dimensional lattice $\Z$ in
order to simplify the notations. For each subset $\Lambda_n$
in the lattice $\Z$ written as
 $ \Lambda _n = \{ -n , ... , +n \}$ ($ n \geq 1$),
we define a Hamiltonian $H_{\Lambda _n}$ in
$\R^{\Lambda_n}$ by:
%----
$$H_{\Lambda _n}= -{1 \over 2} \sum _{\lambda \in \Lambda _n}
{\partial ^2 \over \partial x_{\lambda}^2}  \ +\ V_{\Lambda_n},
\hskip 1cm  V_{\Lambda_n}  = V_{\Lambda_n}^{quad}
+ V_{\Lambda_n}^{pert}. \leqno (1.1)$$
%-------
where the potential  $ V_{\Lambda_n}^{quad}$ is a definite positive
quadratic form on $\R^{\Lambda _n}$ and where
 $V_{\Lambda_n}^{pert}$ is viewed as a perturbation of
$V_{\Lambda_n}^{quad}$.

\bigskip
The quadratic potential is defined for all $n$ by:

%---
$$  V_{\Lambda_n}^{quad} (x)=  { a \over 2} |x|^2 - b  \sum
_{\lambda = -n}^{n-1} x_{\lambda } x_{\lambda +1} \leqno (1.2)$$
%----
where $a$ and  $b$ are two real numbers verifying $a>2b>0$.
\bigskip

Precise hypotheses on the  perturbation  potential are stated in $(H_1)$
and $(H_2)$ (see below). These assumptions imply that
$V_{\Lambda_n}^{pert}$ is a multiplication operator by a real-valued
function
$v_{\Lambda_n}^{pert}$ belonging to $C^3(\R^{\Lambda_n})$, and satisfying
$v_{\Lambda_n}^{pert}(x) = o (|x|^2)$ near infinity.
\bigskip
Following Kato-Rellich's theorem, the operator $H_{\Lambda _n}$ defined in
(1.1), with the hypotheses $(H_1)$ and $(H_2)$, is self-adjoint with the
same domain as the harmonic oscillator on  $\R^{\Lambda_n}$. Hence, we
can  define the  unitary operator
$e^{it H_{\Lambda _n}}$ ($t\in \R$).

 \bigskip

Thus,
 the following operator  is well-defined:
 %------
 $$ \alpha _{\Lambda _n}^{(t)} (A) = e^{it H_{\Lambda _n}}
 A e^{-it H_{\Lambda _n}} \leqno (1.3)$$
for all $A \in {\cal L}({\cal H}_{\Lambda _n})$, (where ${\cal
H}_{\Lambda _n} = L^2(\R^{\Lambda _n})$), and for all $t\in \R$.
 %----
 It is then natural to ask whether this sequence of operators has a limit when
  $n$ tends to $+\infty$, and for which class of operators $A$ ?
 More precisely, we are
looking for a Banach algebra ${\cal A}$ satisfying the following
conditions:
\medskip
 - The spaces ${\cal L} (L^2(\R^{\Lambda}))$, (where $\Lambda$
 is a finite subset of $\Z$), is isometrically immersed in the algebra (the
 elements of ${\cal L} ({\cal H}_{\Lambda})$ are under this identification
called local observables
 supported in $\Lambda$).
 \medskip
 - For all local observables $A$, the limit as $n$ tends to infinity of
$\alpha _{\Lambda _n}^{(t)}
 (A)$, denoted by $\alpha ^{(t)} (A)$, exists in this algebra ${\cal A}$.
\medskip
 - This operator $\alpha^{(t)}$, defined in this procedure for local
observables $A$,
 may be extended by density to the whole algebra  ${\cal A}$,  and acts in
a continuous way.
 \bigskip

 Several works, related to this issue, have considered the
 $C^{\star}-$algebra ${\cal A}$ of the quasi-local observables. Let us
recall its definition (cf [SI]).  For each finite subset $\Lambda$ in $\Z$
 set ${\cal H}_{\Lambda } = L^2(\R^{\Lambda })$. One notes that,
 if $\Lambda \subset \Lambda' $ then ${\cal L} ({\cal H}_{\Lambda } )$
is isometrically immersed in ${\cal L} ({\cal H}_{\Lambda' } )$.
 Therefore, one may define  ${\cal A}$ as the completion of the inductive
limit of the spaces ${\cal L} ({\cal H}_{\Lambda } )$:
 %---
 $$ {\cal A} = \overline { \bigcup _{\Lambda \subset \Z}
 {\cal L} ({\cal H}_{\Lambda } ) } \leqno (1.4)$$
 %---
 \bigskip
This algebra is well-adapted in the case of bounded potentials, or when the
first order derivatives are bounded
  (cf  e.g. the work of
[NOS] for the existence of a dynamics,  or [ACLN] for estimates on the decay
of the correlations),  whereas it might not be suitable for the perturbated
quadratic case studied here.

\medskip
Another algebra,  {\it the  Weyl algebra}, is considered by
Malyshev-Minlos [MA-MI] and by Thirring [TH]. This algebra fits to the
non-perturbated quadratic case ($
V_{\Lambda_n}^{pert}=0$),  and it is defined using the Fock space's
formalism.
\bigskip
The space  ${\cal H}$ denotes the symmetrized Fock space ${\cal H} =
F_s (\ell^2(\Z))$, associated to  the Hilbert space $\ell^2(\Z)$.
For all $\lambda \in \Z$, on defines the two self-adjoint operators
$P_{\lambda}$ and  $Q_{\lambda}$ in the Fock space, verifying the same
commutation relations as the   position and momentum operators in
$L^2(\R^n)$. (Note that there is here an infinite number of these
operators.)   For each finite subset $\Lambda$ de $\Z$, the space ${\cal L}
({\cal H}_{\Lambda }
)$ (where ${\cal H}_{\Lambda}= L^2(\R^{\Lambda})$) is isometrically
immersed in ${\cal L} ({\cal H})$. This identification
extends also to non bounded operators. Thus,  the  multiplication operator by
$x_{\lambda}$ and the operator
${1 \over i} {\partial \over \partial x_{\lambda}}$ ($\lambda \in
\Lambda $) becomes the two operators $Q_{\lambda}$ and
$P_{\lambda}$, sometimes denoted in this paper by
$Q_{\lambda}^{(0)}$ and $Q_{\lambda}^{(1)}$:
%----
$$ Q_{\lambda}^{(0)} = Q_{\lambda} = x_{\lambda} \hskip 1cm
Q_{\lambda}^{(1)} = P_{\lambda} = {1 \over i} {\partial \over
\partial x_{\lambda}} \leqno (1.5) $$
%----
\bigskip
The Fock spaces formalism  allows us to properly define, for all real
sequences $u$ and $v$ in $\ell^2(\Z)$,
the non bounded self-adjoint operator, (the Segal operator), formally
defined by:
%----
$$\Pi (u, v) = \sum _{\lambda \in \Z} (u_{\lambda } P_{\lambda} +
v_{\lambda } Q_{\lambda}) \leqno (1.6)$$
%---
The two operators $P_{\lambda}$ and $Q_{\lambda}$ are generally not
defined by (1.5) anymore, but, instead, $\Pi (u, v)$ is defined starting from
the  creation and annihilation operators associated to $\ell^2(\Z)$ (see
section 2). The corresponding unitary operator $W(u , v)=
e^{i \Pi (u , v)}$ is called a
 Weyl operator.
\bigskip
The Weyl algebra introduced by Malyshev-Minlos [MA-MI] or by Thirring [TH]
is the closure in ${\cal L}( {\cal H})$ of the
subspace generated by the operators $W(u , v)$ ($u$ and $v$
being real sequences in $\ell^2(\Z)$).

\bigskip

In the purely quadradic case ($V_{\Lambda_n}^{pert} = 0$) and for all $A$
in this Weyl agebra, an explicit analysis
allows us to define properly $\alpha^{(t)}_{\Lambda_n}(A)$ (even if $A$ is
not supported in $\Lambda_n$) and to define the limit
operator  $\alpha^{(t)}(A)$ such that, for all $f\in
{\cal H}$:
%-----
$$ \lim _{n\rightarrow \infty} \Big \Vert\ [
\alpha^{(t)}_{\Lambda_n}(A) - \alpha^{(t)}(A)] f \Big \Vert _{\cal
H} \ =\ 0$$
%-----
In order to derive the latter limit, uniform estimates, such as those
established  in [N-R-S-S], are needed.

\bigskip

Using the Weyl algebra defined above, it is probably difficult to also
obtain  these results when the potential of perturbation is turned on. The
purpose of this work is then to extend the above results to the quadratic
case with perturbations by involving another algebra ${\cal W}_2$
included in ${\cal L}({\cal H})$. Furthermore, the Lieb-Robinson estimates
in [N-R-S-S] are also extended to that framework.

\bigskip
Before giving the definition of ${\cal W}_2$, let us mention that the works
of Calderon-Vaillancourt [C-V] and Beals [BE] (see also H\"ormander [HO]),
give an important role to a particular subalgebra of ${\cal L}(L^2(\R^n))$
or here, of
${\cal L}(L^2(\R^{\Lambda}))$,  for all finite subset $\Lambda$
in $\Z$. This particular subalgebra $OPS^0 (\R^{\Lambda})$ is the set  of
pseudo-differential
operators on $\R^{\Lambda}$,
associated to symbols that are bounded, together with all of their derivatives. From
Beals [BE],  these operators are characterized by the following
property, implying the operators $Q_{\lambda }^{(0)}$
 and $Q_{\lambda }^{(1)}$ defined in (1.5) for all
$\lambda \in \Lambda$. An operator $A$ in ${\cal
L}(L^2(\R^{\Lambda}))$ is in $OPS^0 (\R^{\Lambda})$ if, and only if, all the
iterated commutators $(ad \ Q_{\lambda
_1}^{k_1} ) \ ... (ad \ Q_{\lambda _m}^{k_m} ) A$, (with $\lambda
_1$, ... $\lambda _m$ are in $\Lambda$, $m\geq 0$, and $k_j\in
\{ 0, 1 \}$),  are bounded operators in
$L^2(\R^{\Lambda})$. (The commutators are known to a priori map from ${\cal
S}(\R^{\Lambda})$ into ${\cal S}'(\R^{\Lambda})$.)
\bigskip
Replacing $\Lambda$ by $\Z$, one may analogously define a
decreasing sequence of subalgebras ${\cal
W}_k$ in  ${\cal L}({\cal H})$ ($k\geq 0$). Set ${\cal W}_0 =
{\cal L}({\cal H})$. We denote by ${\cal W}_1$ the set of all $A$ in
${\cal W}_0$ such that, for all $\lambda \in \Z$, the
commutators $[A , Q_{\lambda}]$ and $[A , P_{\lambda}]$ are bounded in
${\cal H}$, and such that the sum in the following norm is finite:
%-----
$$ \Vert A \Vert _{{\cal W}_1} =  \Vert A \Vert _{{\cal W}_0} +
\sum _{\lambda \in \Z \atop k= 0, 1}  \Vert \ [  A ,
Q_{\lambda}^{(k)}]\
 \Vert _{{\cal W}_0} \leqno (1.7)$$
 %----
Note that the above commutators are properly defined in section 2.
From now on, the operators $Q_{\lambda}^{(0)}= Q_{\lambda}$ and
 $Q_{\lambda}^{(1)}= P_{\lambda}$ are defined through the Fock space's
formalism, and not  by (1.5) anymore.

\medskip
Let us denote by ${\cal W} _2$ the set of all
operators $A \in {\cal W} _1$  such that  the commutators $[
Q_{\lambda}^{(k)}, A ]$
  belongs to ${\cal W} _1$  for all $\lambda $
in $\Z$, and such that
the sum in the norm below is finite:
%-----
$$ \Vert A \Vert _{{\cal W}_2} =  \Vert A \Vert _{{\cal W}_1} + {1
\over 2} \sum _{(\lambda, \mu) \in \Z^2 \atop 0 \leq j, k\leq 1}
\Big \Vert \ \big [[ A , Q_{\lambda}^{(j)}] , \ Q_{\mu}^{(k)}\big
] \
 \Big \Vert _{{\cal L}({\cal H})} \leqno (1.8)$$
 %----

\bigskip

\noindent {\it An example.} For all $u$ and  $v$ in $\ell^1 (\Z)$,
the Weyl operator $ W(u , v) = e^{i \Pi  (u , v)}$ is in
${\cal W}_k$ ($0 \leq k \leq 2$).

\bigskip
One might define similarly a sequence of algebras ${\cal W}_k$ using
iterated commutations.
In particular, the intersection set of these algebras could correspond to
an analogous of $OPS^0$ in infinite dimension. Other particular classes of
pseudo-differential operators in
infinite dimension are studied by B. Lascar (see  [L1]  [L2],...).

\bigskip
Among all of these algebras and for our point of view,  it is ${\cal W}_2$
that appears to be the most suitable to our study.
If $A$ is not supposed to be  an element of ${\cal W}_2$,  $A$ being only
assumed to be in ${\cal L} ({\cal H})$ and  supported on a finite subset
$E$ of $\Z$, it appears to be possible to show that, for all $f$ in ${\cal
H}$, the sequence $\alpha _{\Lambda
_n}^{(t)}(A)f$ weakly converges in ${\cal H}$.  If this limit is denoted by
$\alpha ^{(t)}(A)f$, it is not clear  whether the map  $t \rightarrow
\alpha^{(t)}$ is continuous, neither whether $\alpha^{(t)}$ may be extended to
a suitable Banach algebra.

\bigskip
More precise estimates are obtained when the local observable $A$ belongs
to  ${\cal W}_2$. Before that, let us describe now  the
perturbation potential.

\bigskip
\noindent {\it Hypotheses on the perturbation potentials.}
The operator $V_{\Lambda_n}^{pert}$ is written as the following sum:
%----
$$ V_{\Lambda_n}^{pert} = \sum _{ \lambda \in \Lambda_n}
V_{\lambda} + \sum _{(\lambda , \mu)\in \Lambda _n^2 \atop \lambda
\not=\mu }
 V_{\lambda \mu},  \leqno (1.9)$$
%--------
where the  operators $V_{\lambda}$ and $V_{\lambda \mu}$ are
defined for all $\lambda$ and $\mu$ in $\Z$,  and verify the assumptions below:
\medskip
\noindent (H1) For each pair $(\lambda , \mu)$ of $\Z$ with $\lambda \not
= \mu$, $V_{\lambda \mu}$ is a multiplication by a $C^3$ real-valued
function $v_{\lambda \mu}$  depending  only on the variables $x_{\lambda}$
and
$x_{\mu}$. Moreover, denoting $\widehat {v_{\lambda \mu} }$ the Fourier
transform of $v_{\lambda \mu}$  (on  $\R^2$ and in the sense of
distributions), the distributions $\xi_{\lambda
}^j \xi_{\mu}^k \widehat {v_{\lambda \mu} }$
belongs to $L^1(\R^2)$ if $2 \leq j+k \leq 3$. Furthermore, there exists
$ C_0>0$ and $\gamma_0>0$, (not depending on $\lambda$ and $\mu$),
such that:
%----
$$ \sum _{2 \leq j+k \leq 3 } \Vert\xi_{\lambda }^j
 \xi_{\mu}^k \widehat {v_{\lambda \mu} }\Vert _{L^1(\R^2)} \ \leq
 C_0 e^{-\gamma_0 |\lambda -\mu|}, \leqno (1.10)$$
 %-----------
 $$ | \nabla v_{\lambda \mu} (0) | \leq  C_0 e^{-\gamma_0 |\lambda
 -\mu|}. \leqno (1.11)$$
 %----
\medskip \noindent (H2) For each point $\lambda$ in $\Z$,
$V_{\lambda}$ is the multiplication by a $C^3$ real-valued function
$v_{\lambda}$
 depending only on the
 variable $x_{\lambda}$. If we denote by $\widehat {v_{\lambda}}$ the
 Fourier transform of $v_{\lambda}$, the distributions $\xi _{\lambda}^j
\widehat {v_{\lambda}}$ are in $L^1(\R)$ when $2 \leq j \leq 3$,
and
%----
$$ \sum _{2 \leq j \leq 3} \Vert \xi_{\lambda}^j \widehat
{v_{\lambda}} \Vert _{L^1(\R)} \leq C_0, \hskip 1cm
 | \nabla v_{\lambda} (0) | \leq  C_0. \leqno (1.12)$$
%-----

\bigskip
In particular, in the case of  {\it interactions between nearest
neighbors,
} one has  $V_{\lambda \mu} =0$ if $|\lambda -
    \mu| \geq 2$. It is then sufficient that the integrals in the l.h.s.
of (1.10) and (1.12) are uniformly  bounded in $\lambda$. In that case,
the hypotheses (H1)
    are (H2) satisfied for any  $\gamma_0>0$ and in all the results below,
    the phrase
\og for all $\gamma \in
    ]0, \gamma_0[$\fg  is replaced by \og for all
    $\gamma>0$\fg.
\bigskip
For each integer $n$, the  perturbation potential   $V_{\Lambda_n}^{pert}$
and the  Hamiltonian $H_{\Lambda _n}$ are defined
by   (1.9)
 and (1.1) respectively. In [NRSS], the authors have only considered the
$V_{\lambda}$'s. We shall say that an
element $A$ of ${\cal W}_2$ has a finite support if there exists a finite
subset
 $E$ in $\Z$, such that $A$ is identified to an element of  ${\cal L}
({\cal H}_E)$. The smallest set having this property is called the support
of $A$ and is denoted by
$\sigma (A)$.

\bigskip
\noindent {\bf Theorem 1.1.}  {\it Under the above hypotheses,  for all
element $A \in {\cal W} _2$ with finite support, for all $t\in \R$, for
all $n$ such that $\Lambda_n$
 contains the support of $A$, the operator $\alpha _{\Lambda _n}^{(t)} (A)$
belongs to ${\cal W}_2$. Moreover, there exists two real positive real
numbers $C$ and
 $M$ not depending on $n$ and $t$ such that:
 %------
$$ \Vert \alpha _{\Lambda _n}^{(t)} (A) \Vert _{{\cal W}_2} \leq
 C e^{M|t|} \Vert A \Vert_{{\cal W}_2}. \leqno (1.13)$$
 %--------
Furthermore, for each $f\in {\cal H}$, the sequence $\alpha _{\Lambda
_n}^{(t)}
 (A)f$ strongly converges in ${\cal H}$. Denoting this limit by
$\alpha^{(t)} (A)f$, the map $t \rightarrow \alpha^{(t)} (A)f$ is strongly
continuous, the operator $\alpha ^{(t)}
 (A)$ is in ${\cal W}_2$ and one has:
 %------
$$
 \Vert \alpha ^{(t)} (A) \Vert _{{\cal W}_2} \leq
 C e^{M|t|} \Vert A \Vert_{{\cal W}_2}. \leqno (1.14)$$
 %-----
 }
 \bigskip
In the first part of this theorem,  (where $n$ is fixed),
one may think that $ \alpha _{\Lambda _n}^{(t)}$ acts in the algebra
${\cal W}_k$, defined similarly as ${\cal W}_1$ and ${\cal
W}_2$, but with iterated commutators of length $k$, and for
operators supported in $\Lambda_n$. (The hypotheses (H1) and (H2) naturally
need to be strengthened.)  From Beals characterization, one would deduce a
group action of $ \alpha _{\Lambda _n}^{(t)} $ on the operators in
$OPS^0(\R^{\Lambda})$.  An alternative approach concerning this problem
may be found in the works of Bony (see [BO1] and [BO2]).

\bigskip
Moreover, under the hypotheses of  theorem 1.1, the
  automorphism $\alpha^{(t)}$, (initially defined for local observables), is
extended in a unique way to the whole
  algebra ${\cal W}_2$ (see below). To this end, we introduce
    Sobolev-type spaces.

\bigskip
%-----
Let ${\cal H}^2$ be the subspace of the $f\in {\cal H} $ such that the
following norm is finite:
%----
$$ \Vert f \Vert _{{\cal H}^2} \ =\  \Vert f \Vert _{{\cal H}} + \
\sup _{\lambda  \in \Z \atop 0 \leq j \leq 1 } \Vert
Q_{\lambda}^{(j)} f \Vert _{\cal H}\ +\ \sup _{(\lambda , \mu )
\in \Z^2 \atop 0 \leq j , k\leq 1} \Vert Q_{\lambda }^{(j)}
 Q_{\mu } ^{(k)} f \Vert _{\cal H}.
\leqno (1.15)$$
%------------
\bigskip
 Since a convergence in norm is needed,  theorem 1.1 is now completed with
the result below:

\bigskip
\noindent {\bf Theorem 1.2.}  {\it There exists $C>0$, $\gamma
>0$ and $M>0$ with the following properties. For all  $A$
in ${\cal W}_2$ with a finite support denoted by $\sigma (A)$,  for all $n$
such that $\Lambda_n$ contains $\sigma (A)$ and for all
$t\in \R$, we have:
%-----
$$ \left \Vert  \Big [\alpha ^{(t)} _{\Lambda _n}(A) - \alpha
^{(t)}(A) \Big ]  \right \Vert _{{\cal L}({\cal H}^2 ,{\cal H})}
\leq C e^{M|t|} e^{-\gamma d( \sigma (A), \Lambda_n^c)}  \Vert A
\Vert _{{\cal W}_2} \leqno (1.16)$$
%---
Moreover,
%---
$$
 \Vert \alpha ^{(t)} (A) \Vert _{{\cal L}({\cal H}^2 , {\cal H})} \leq
 C e^{M|t|}  \Vert A \Vert _{{\cal L}({\cal H}^2 , {\cal H})} \leqno
(1.17)$$
 %---
}
%----
\bigskip

The set of all observables having a finite support is not dense in
${\cal W}_2$. In order to extend $\alpha ^{(t)} $,  we shall use,
instead of density, the following two results.

\bigskip
 \noindent {\bf Theorem 1.3.} {\it  Set $A$ in ${\cal W}_2$.
 Then there is a sequence $(A_n)$ in ${\cal W}_2$ such that each
 $A_n$ has a finite support, and such that:
 %----
 $$ \Vert A_n \Vert _{{\cal W}_2 } \leq \Vert A \Vert _{{\cal W}_2 },
\hskip 1cm  \lim  _{n \rightarrow \infty }  \Vert A_n -A  \Vert
_{{\cal L}({\cal H}^2 , {\cal H} )} = 0. \leqno (1.18)$$
%------
}
%------
\bigskip
\noindent {\bf Theorem 1.4.} {\it Let $(A_n)$ be a sequence of operators
in ${\cal W}_2$. Suppose that $\Vert A_n \Vert
_{{\cal W}_2} \leq 1$ and assume that there exists $A\in {\cal L} ({\cal
H}^2,
{\cal H})$ such that $\Vert A_n - A \Vert _{{\cal L} ({\cal H}^2 ,
{\cal H})}$ tends to $0$. Then $A$ may be extended to an element of ${\cal
L}({\cal H})$ which belongs to ${\cal W}_2$ and
 $\Vert A \Vert _{{\cal W}_2} \leq 1$. Moreover, for all $f\in
 {\cal H}$ the sequence $A_nf$ converges to
 $Af$ in ${\cal H}$.
}
%------

\bigskip
Consequently, we easily deduce from theorems 1.1 - 1.4 that $\alpha^{(t)}$
may be extended, in a unique way, to the whole algebra ${\cal
W}_2$, without any conditions on the finiteness of the supports (see
section 7).
The map $\alpha^{(t)}$ is not a ${\cal
W}_2$ norm conservative map, but it is ${\cal L}({\cal H})$ norm
conservative. Using this point,
 $\alpha^{(t)}$ is extended to the closure $\overline {{\cal
W}_2}$ of ${\cal W}_2$ in ${\cal L}({\cal H})$. Thus, $\alpha^{(t)}$ acts
in $\overline {{\cal W}_2}$ in a continuous way (for the simple topology) and is
norm conservative.

\bigskip
\noindent {\it  Lieb-Robinson's inequalities.}
\medskip
These inequalities, established in [L-R] for bounded
Hamiltonians and, more recently, in [N-R-S-S] for quadratic Hamiltonians,
express the propagation of the correlation between two
observables with separated supports, as a function of the time and of the
distance between the two supports.
\medskip
For all $h$ in $\Z$,  set $T_h$ the map in $\ell^2
(\Z)$ defined by $(T_hu )_{\lambda} = u_{\lambda + h}$ for all $u\in
\ell^2(\Z)$ and for all $\lambda \in \Z$. With $T_h$ we define a map in
the Fock space ${\cal
H} = F_s(\ell^2(\Z))$ that is still noted $T_h$. For any $A$ in ${\cal
L}({\cal H})$ we set $\tau_h(A) = T_h^{-1} A T_h$.
\bigskip
In our framework, the Lieb-Robinson type inequalities have the
following form:
\bigskip
\noindent {\bf Theorem 1.5.}  {\it There exists a real number
$v_0$ with the following property. For any elements $A$
and $B$ of ${\cal W}_2$ with finite supports, for any sequence $(h_n
, t_n)$ tending to infinity in $\Z \times \R$ and satisfying
$|h_n| \geq v_0 | t_n|$, for any $f\in {\cal H}$, we have:
%----
$$ \lim _{n \rightarrow \infty} \big [ \alpha^{(t_n)} (A) \ , \
\tau _{h_{n}} (B) \big ] f \ =\ 0.  \leqno (1.19)$$
%-----
}

\bigskip
The infimum $V_0$, of the all the $v_0$ satisfying the above property,
defines a kind of propagation speed, which is different from the usual
definitions of phase  and group velocities (cf Cohen-Tannoudji [C-T]).
\medskip
In the case of cyclic quadratic potentials, (that is to say, without any
perturbation,  but obtained by  adding to $V_{\Lambda_n}^{quad}$ of
(1.2)  an end point interaction potential
 $- b x_nx_{-n}$, one finds in
[N-R-S-S] an estimate of this propagation speed. (In [NRSS] this is written
for a  multidimensional lattice model.)  We shall provide here an alternative
estimate of the same type, with an elementary proof, given in section 4.
The analysis of chains of harmonic oscillators with  cyclic interactions usually involves the
dispersion relation
 $\omega
(\theta) = \sqrt {a -2b \cos \theta }$,  (cf [C-T]). It is then natural to
define a complex version of this relation,  and to define:
%----
$$\Omega (z) = \sqrt {a  - b(z+ z^{-1})}, \hskip 1cm z\in {\bf
C}\setminus \{ 0\}. $$
%-----
For any $\gamma >0$, set:
%---
$$ M(\gamma) = \sup _{|z|= e^{\gamma}} | {\rm Im} \ \Omega (z)|. $$
%----
The propagation speed verifies, in the cyclic quadratic case:
%----
$$ V_0 \leq \inf _{\gamma >0} {M(\gamma)\over \gamma} . $$
%-----
In a more general case, this estimate is less precise. For all $\gamma$ in
$]0, \gamma_0[$ ( $\gamma_0$ being the real number appearing in the
hypotheses
(H1) and (H2)), we shall define in Proposition 3.4 a real number
$S_{\gamma}$ and we shall prove in  section 8 that the propagation speed
verifies :
%----
$$  V_0 \leq  \inf _{0  < \gamma < \gamma _0} {2 \sqrt {
S_{\gamma}} \over \gamma}. $$
%-----
The constant number $S_{\gamma}$ depends only on $a$ and $b$,  together with
the
norms in ${\cal F} L^1 (\R)$ or ${\cal F} L^1 (\R^2)$ of the second
derivatives of the potentials of perturbation. We then note that,
multiplying
$a$, $b$ and the potentials of perturbation  by a constant
$g>0$, our estimates on the propagation speed is multiplied by $\sqrt {g}$.

\bigskip

Section 2 is concerned with the subalgebra  ${\cal W}_k$. In section 3,
properties  on $V_{\Lambda_n}$ under the hypotheses (H1) and (H2)
are established.  Evolution operators, for finite systems on the lattice,
are studied in  sections 4 - 6. Sections 7 and 8 are respectively devoted
to
perform the limit $n$ goes to infinity (the number of sites tends to
infinity), and to derive the Lieb-Robinson's inequalities.
\bigskip

We are grateful to M. Khodja for helpful discussions.

\bigskip

%XXXXXX-SECTION-2-XXXXXXXXXXXXXXX

\noindent {\bf 2. Algebras of operators in the Fock space. }

\bigskip
\noindent {\it Notations  on the Fock spaces (cf [RE-SI]).}
\medskip
For any $E$ subset of $\Z$, the symmetrized Fock space
associated to the Hilbert space  $\ell^2(E)$ shall be denoted ${\cal
H}_E$. When $E= \Z$, this space is still noted ${\cal H}$.
The ground state of ${\cal H}_E$ is denoted by
$\Omega_E$ or $\Omega$ when $E= \Z$.

\bigskip
If $E_1$ and $E_2$ are two disjoint sets of $\Z$ one may identify ${\cal
H}_{E_1 \cup E_2} $ and ${\cal H}_{E_1} \otimes
{\cal H}_{E_2}$ (the completed tensorial product). On may also identify
$\Omega _{E_1 \cup E_2} $ with $\Omega _{E_1} \otimes
\Omega _{E_2}$.
\bigskip
For all real sequence $u$ in $\ell^2(\Z)$ we define the two non bounded
operators $a(u)$ (annihilation operator) and
$a^{\star }(u)$ (creation operator), being each other the formal
adjoint,
and verifying the following commutation relations:
%------
$$ [ a(u) , a(v) ] =  [ a^{\star}(u) , a^{\star} (v) ]= 0 \hskip
1cm [ a(u) , a^{\star}(v) ] = (u , v), $$
%----
for all $u$ and $v$ in $\ell^2(\Z)$.
\bigskip
We shall denote by $(e_{\lambda})_{(\lambda \in \Z)}$ the canonical basis
of
$\ell^2(\Z)$.  Starting from the ground state $\Omega$, and applying
successively the creation operators, one defines $a^{\star}(e_{\lambda_1})
... a^{\star}(e_{\lambda_m})\Omega$,  being orthogonal elements of ${\cal
H}$.  Let  ${\cal D}$ be the subspace of ${\cal H}$ generated by these
vectors. It is known that
${\cal D}$ is dense in ${\cal H}$. The space ${\cal D}$ is included in the
domain of all  $a(u)$ and $a^{\star}(u)$,
($u\in \ell^2(\Z)$). For all $f$ in ${\cal D}$ there exists a finite
subset $S \subset \Z$ such that $f$ is written as: $f = g
\otimes \Omega _{S^c}$ with $g\in {\cal H}_S$. We then say that $f$
is supported in $S$.
\bigskip
Next we define the Segal operator $\Pi (u , v)$ by:
%-----
$$\Pi   (u , v ) = { a(u) + a^{\star} (u) \over \sqrt {2}}  \ + \
{ a(v) - a^{\star} (v) \over i\sqrt {2}} \leqno (2.1)$$
%----
for all real elements $u$ and $v$
in $\ell^2(\Z)$.
An element $f\in {\cal H}$ is the domain of $\Pi   (u ,
v )$ if there exists a sequence $(f_n)$ in ${\cal D}$ such that $f_n$
converges to $f$ in ${\cal H}$ and such that $\Pi   (u , v ) f_n$
has a limit in ${\cal H}$. Thus, $\Pi   (u , v )$ is a self-adjoint
operator. The associated Weyl operator is $W(u
, v) = e^{i \Pi (u , v)}$.
\bigskip
In particular, for each element $e_{\lambda}$  in the canonical basis of
$\ell^2(\Z)$ the Segal operators are noted:
%----
$$Q_{\lambda} = Q_{\lambda}^{(0)} = { a(e_{\lambda}) + a^{\star}
(e_{\lambda}) \over \sqrt {2}} \hskip 1cm P_{\lambda} =
Q_{\lambda}^{(1)} = { a(e_{\lambda}) - a^{\star} (e_{\lambda})
\over i\sqrt {2}}. \leqno (2.2)$$
%----
\bigskip
Let us write down an orthonormal basis. We shall limit ourselves to the
Hilbert space ${\cal H} _{ \{ \lambda \} }$
associated  to a subset of $\Z$ reduced to one element $\lambda$. In this
space we again used the
construction of ${\cal D}$ and obtain the basis
$(h_n)_{(n\geq 0)}$ being now normalized by setting:
%----
$$ h_0 = \Omega _{ \{ \lambda \} } \hskip 1cm h_{j+1} =
(j+1)^{-1/2} a^{\star} (e_{\lambda}) h_j \hskip 1cm (j\geq
0)\leqno (2.3)$$
%----
The space ${\cal H} _{ \{ \lambda \} }$ may be identified with $L^2(\R)$
in an isometric way.  Then the basis $(h_j)$ becomes the Hermite's
functions basis, and  the operators
$Q_{\lambda} $ and $P_{\lambda}$ respectively become the multiplication
by $x_{\lambda}$ and the operator ${1 \over i}
{\partial \over \partial _{x_{\lambda}}}$.  Effectuating the completed
tensorial product, the space  ${\cal
H}_{\Lambda}$ is similarly identified to $L^2(\R^{\Lambda})$ for each
finite subset $\Lambda$ of $\Z$.

\bigskip
For any $E \subset F \subseteq \Z$, and any operator $T
\in {\cal L}(E)$, we define $i_{EF} (T)$ by the following equality:
%----
$$ i_{EF} (T) \ =\  T \otimes I_{F \setminus E}, \leqno (2.4)$$
%----
where $I_{F \setminus E}$ is the identity in the space  ${\cal H}_{F
\setminus
E}$. In particular, if $F = \Z$ the operator $i_{E
\Z} (T)$ is said to be supported in $E$.

\bigskip

\bigskip

\bigskip
\noindent {\it   Sobolev spaces.}  Let us denote by ${\cal H}^1$
the set of all $f\in {\cal H}$ such that
$f$ belongs to the domains of the Segal operators $Q_{\lambda}
= Q_{\lambda}^{(0)}$ and  $P_{\lambda} = Q_{\lambda}^{(1)}$ for all
$\lambda \in \Z$,
 and such that the following norm is finite:
%---
$$ \Vert f \Vert _{{\cal H}^1} = \Vert f \Vert_{{\cal H}} + \sup
_{\lambda \in \Z \atop 0 \leq j \leq 1} \   \Vert
Q_{\lambda}^{(j)} f \Vert_{{\cal H}}   \leqno (2.5)$$
%-----
The space ${\cal H}^2$ is the set of all $f\in {\cal H}^1$ such that
$Q_{\lambda}^{(0)}f$ and
 $Q_{\lambda}^{(1)}f$
belongs to ${\cal H}^1$ for all $\lambda$ in $\Z$,  and with a finite
following norm:
%---
$$ \Vert f \Vert _{{\cal H}^2} = \Vert f \Vert_{{\cal H}^1} + \sup
_{(\lambda , \mu) \in \Z^2 \atop 0 \leq j, k \leq 1} \ \Vert
Q_{\lambda}^{(j)} Q_{\mu}^{(k)}f \Vert_{{\cal H}}   \leqno (2.6)$$
%-----
These spaces are dense in ${\cal H}$ since they contain
${\cal D}$. If $E$ is a subset of $\Z$ then the subspace ${\cal H}^k_E$ is
defined analogously in its corresponding Hilbert space
 ${\cal H}_E$.

\bigskip
\noindent {\it Commutators, and  spaces with negative orders.}

\medskip
For all $A$ in ${\cal L}({\cal H})$, for all $f\in {\cal
H}^1$ and for any $\lambda \in \Z$ the map:
%---
$$ {\cal H}^1 \ni g \rightarrow \big < A Q_{\lambda}^{(j)} f \ ,\ g \big >
\ - \ \big < A
 f \ ,\ Q_{\lambda}^{(j)}   g \big >  \hskip 2cm 0 \leq j \leq 1 \leqno
(2.7)$$
 %----
 is a continuous antilinear map on the space ${\cal H}^1$.
 We denote by  ${\cal H}^{-k}$ the anti-dual of ${\cal H}^k$ ($0 \leq k
  \leq 2$). For any $A$ in ${\cal L}({\cal H})$ the map (2.7) is linear
and continuous from ${\cal H}^1$ to ${\cal H}^{-1}$. It is noted $[A ,
Q_{\lambda}^{(j)} ]$.  One may identify ${\cal H}$
with a subspace of ${\cal H}^{-1}$,  and the latter one is identified to a
subspace of
 ${\cal H}^{-2}$. Thus, the operators $
Q_{\lambda}^{(j)}$  are bounded from ${\cal H}^m$ to ${\cal
H}^{m-1}$ ($-1\leq m \leq 2$),  and this allows us to define the
iterated
commutators  $\Big [ Q_{\lambda}^{(j)}  , [
Q_{\mu}^{(k)} , A ] \ \Big ]$, ($(\lambda , \mu) \in \Z^2$, $0
\leq j , k \leq 1$), as continuous linear maps from
${\cal H}^2$ to ${\cal H}^{-2}$. This map is also denoted by $ (ad
Q_{\lambda}^{(j)} )\ (ad Q_{\mu}^{(k)} )\ A$.
 \medskip

If there is real number $C>0$ verifying :
%----
$$ \Big |\  \big < A  Q_{\lambda}^{(j)}  f \ ,\ g \big > \ - \ \big  < A
 f \ ,\  Q_{\lambda}^{(j)} g \big > \ \Big | \ \leq \ C \ \Vert f \Vert
 _{{\cal H}}\  \ \Vert g \Vert _{{\cal H}}$$
 %-----
for all $f$ and $g$ in
${\cal H}^1$
we shall say that the commutators $[A , Q_{\lambda}^{(j)} ]$ are in ${\cal
L}({\cal
H})$. Then, for all $f$ in ${\cal H}^1$ there exists an element
${\cal H}$ noted $[A , Q_{\lambda}^{(j)} ]f$ such that we have:
%--
$$  \big < A P f \ ,\ g \big > \ - \ \big < A
 f \ ,\ Q_{\lambda}^{(j)}  g \big > \ = \ \big < [A , Q_{\lambda}^{(j)} ]f
\ ,\ g \big >$$
 %----
for all $g$ in ${\cal
H}^1$,
and the previously defined operator $[A , Q_{\lambda}^{(j)} ] : {\cal H}^1
\rightarrow {\cal H}$ is extended to an element of ${\cal L}({\cal
 H})$.
Proceeding similarly, one gives a precise meaning to
\og the commutator $\Big [ \ [A , Q_{\lambda}^{(j)} ], Q_{\mu}^{(k)}
\Big ]$ is in ${\cal L} ({\cal H})$\fg

\bigskip
\noindent {\it  Weyl Algebra.}
\medskip

We denote by ${\cal W}_1$ the set of all $A$ in ${\cal L} ({\cal
H})$ having their   commutators $ [
 A ,  Q_{\lambda}^{(j)}]$  ($0 \leq j\leq 1$)  in
 ${\cal L}({\cal H})$ for all $\lambda$ in $\Z$,  and having a finite
following norm:
%----
$$ \Vert A \Vert _{{\cal W}_1} =  \Vert A \Vert _{{\cal L} ({\cal
H})} + \sum _{\lambda \in \Z \atop 0 \leq j \leq 1}  \Vert [ A ,
Q_{\lambda}^{(j)} ] \Vert _{{\cal L} ({\cal H})} \leqno (2.8)$$
%----
We denote by  ${\cal W}_2$ the set of elements $A$ belonging to ${\cal
W}_1$, having  commutators
$\big [ \ [ A , Q_{\lambda}^{(j)} ], Q_{\mu }^{(k)} \big ]$
in ${\cal L} ({\cal H})$ for all $\lambda$ and $\mu$ in $\Z$,  and having a
finite following norm:
%----
$$ \Vert A \Vert _{{\cal W}_2} =  \Vert A \Vert _{{\cal W}_1} \ +
\ {1 \over 2} \sum _{(\lambda, \mu) \in \Z^2 \atop 0 \leq j, k\leq
1} \Big \Vert \ \big [ [A , Q_{\lambda}^{(j)} ],  Q_{\mu}^{(k)}
\big ] \ \Big \Vert _{{\cal L} ({\cal H})} \leqno (2.9)$$
%-------
\bigskip
We easily verify the next proposition.
\bigskip
\noindent {\bf Proposition 2.1.} {\it For all $k\leq 2$ the algebra  ${\cal
W}_k$ is a Banach algebra. For all $A$ and $B$ in
${\cal W}_k$, the following inequality holds:
%----
$$\Vert A B \Vert _{{\cal W}_k} \leq \Vert  A \Vert _{{\cal W}_k}
\Vert  B \Vert _{{\cal W}_k} \leqno (2.10)$$
%-----
Any operator $A \in
{\cal W}_2$ is bounded in the Sobolev spce ${\cal H}^2$ and we have :
%----
$$ \Vert A \Vert _{{\cal L}({\cal H}^2 , {\cal H}^2)} \leq\  3 \
\Vert A \Vert _{{\cal W}_2} \leqno (2.11)$$
%-----
}

%------
\bigskip
\noindent {\it Proof of theorem 1.4.} Let $(A_n)$ be a sequence in ${\cal
W}_2$ and let $A$ be in ${\cal L} ({\cal H}^2 , {\cal H})$
satisfying:
%---
$$ \Vert A_n \Vert _{{\cal W}_2} \leq 1 \hskip 1cm \lim _{n
\rightarrow \infty} \Vert A_n - A \Vert _{ {\cal L} ({\cal H}^2 ,
{\cal H})} = 0$$
%----
For each $f$ in ${\cal H}^2$, one deduces that $\Vert A f
\Vert \leq \Vert f \Vert $ and  $A$ is thus extended by density to an
element of ${\cal L} ( {\cal H})$ with a norm satisfying:
%----
$$ \Vert A \Vert _{{\cal L} ( {\cal H})} \ \leq \liminf
_{n\rightarrow \infty } \Vert A_n \Vert _{{\cal L} ( {\cal H})}$$
%----
For all $\lambda$ in $\Z$, for all $f$ and $g$ in ${\cal
D}$ and for any $n\geq 1$ we see:
%----
$$ \Big |\  \big < A Q_{\lambda}^{(j)}  f \ ,\ g \big > \ - \ < A
 f \ ,\  Q_{\lambda}^{(j)}   g \big > \ \Big | \ \leq \
 \Vert [ A_n , Q_{\lambda}^{(j)}] \Vert  \ \Vert f \Vert
 _{{\cal H}}\  \ \Vert g \Vert _{{\cal H}}  + \varepsilon _n$$
 %-----
where the sequence $\varepsilon_n$ tends to $0$. As a consequence:
%----
$$ \Big |\  \big < AQ_{\lambda}^{(j)}  f \ ,\ g \big > \ - \ < A
 f \ ,\ Q_{\lambda}^{(j)} g \big > \ \Big | \ \leq \ \
  \Vert  \ \Vert f \Vert
 _{{\cal H}}\  \ \Vert g \Vert _{{\cal H}}\ \liminf _{n\rightarrow \infty}
 \Vert [ A_n ,Q_{\lambda}^{(j)}]$$
 %--
Since  ${\cal D}$ is dense in ${\cal H}^1$ this inequality is still valid
for all $f$ and  $g$ in
${\cal H}^1$. With the above definition the commutator $[A ,
Q_{\lambda}^{(j)}]$ is thus in ${\cal
L}({\cal H})$ and one has:
%----
$$ \Vert [A , Q_{\lambda}^{(j)}]\Vert _{{\cal L}({\cal H})} \leq \
\liminf _{n\rightarrow \infty} \Vert [ A_n ,
Q_{\lambda}^{(j)}]\Vert _{{\cal L}({\cal H})}$$
%---
From Fatou's lemma one deduces:
%---
$$ \sum _{\lambda  \in \Z \atop 0 \leq j \leq 1}
\Vert [A ,Q_{\lambda}^{(j)}]\Vert _{{\cal L}({\cal
H})} \leq \liminf _{n\rightarrow \infty} \sum _{\lambda  \in \Z \atop 0
\leq j \leq 1}
 \Vert [A_n , Q_{\lambda}^{(j)}]\Vert _{{\cal L}({\cal H})}$$
 %----
 It is similarly derived that the commutator $[\ [ A , Q_{\lambda}^{(j)}],
Q_{\mu}^{(k)}\ ]$ is in ${\cal
 L}({\cal H})$ for all $\lambda$ and $\mu $ in
 $\Z$ and that:
 %----
$$ \sum _{(\lambda, \mu) \in \Z^2 \atop 0 \leq j, k\leq 1}  \Vert
[\ [A ,Q_{\lambda}^{(j)}], Q_{\mu}^{(k)}\ ]\Vert
_{{\cal L}({\cal H})} \leq \liminf _{n\rightarrow \infty}
 \sum _{(\lambda, \mu) \in \Z^2 \atop 0 \leq j, k\leq 1}  \Vert
[\ [A_n ,Q_{\lambda}^{(j)}], Q_{\mu}^{(k)}\ ]\Vert
_{{\cal L}({\cal H})}$$
 %----
 Theorem 1.4 is then an easy consequence of these points.

 \hfill \carre

%------
\bigskip

In order to derive theorem 1.3,  we shall construct, for each subsets $E$ and $F$ such
that  $E \subset F \subseteq \Z$, an almost right inverse of the
 operator $i _{E, F}$ defined in (2.4).
Set $\Omega _{F \setminus E}$ the ground state  of $F
\setminus E$. Let $\pi _{EF}: {\cal H}_E \rightarrow {\cal
H}_F$ be the map
%----
$$ f \rightarrow \pi _{EF} (f)\ =\ f \otimes \Omega _{F \setminus E},
\leqno
(2.12)$$
%----
 and let $ \pi_{EF}^{\star } $ be the adjoint operator
 $\pi_{EF}^{\star} : {\cal H}_F \rightarrow {\cal H}_E$. Note that
  $ \pi_{EF} ^{\star} \pi _{EF} = I$. For all $A$ in ${\cal
 L} ({\cal H}_F)$ one defines an operator $\rho _{F,
 E}(A)$ in ${\cal L} ({\cal H}_E)$ by:
 %---
 $$ \rho_{F , E}  (A) f = \pi_{EF}^{\star} \circ    A \circ \pi_{EF}
\leqno (2.13) $$
 %-----
\bigskip
Thus, an element   $ \rho_{F , E}  (A) $
 of ${\cal L} ({\cal H}_E)$ is constructed.
One can easily see that, for each $A\in {\cal W}_2$:
%----
 $$ \Vert \rho _{F, E}(A) \Vert _{{\cal W}_2} \leq \Vert A
 \Vert _{{\cal W}_2} \leqno (2.14) $$
 %------
We have also, if $E \subset F \subset G$:
%------
$$ \rho _{GE} = \rho _{FE} \circ \rho _{GF}. \leqno (2.15)$$
%------

 \bigskip
We shall study how an
 operator $A \in
 {\cal L} ( {\cal H}_F)$ may be approximated by
 $i_{EF}\   \rho _{F , E} (A)$ when $E$ is a subset of $F$, being itself finite.

\bigskip
 \noindent {\bf Proposition  2.5.} {\it There exists a real number $C>0$
 such that, for all finite subsets $E$ and $F$  of $\Z$ with
  $E \subset F$, and for all
 $A$ in ${\cal W}_2 $, supported in $F$, one has
%------
 $$\Vert A - i_{EF} \  \rho _{F, E} (A)\Vert _{{\cal L} ({\cal H}^2,
 {\cal H} )} \leq C \sum _{\lambda \in F \setminus E \atop  1 \leq j+k
\leq 2 }
 \Vert (ad P_{\lambda})^j (ad Q_{\lambda})^k A \Vert _{{\cal L}({\cal H})}
 \leqno (2.16)  $$
 %--------
}

\bigskip

This proposition is proven in  Appendix A. Let us show how this
proposition implies  theorem 1.3.

\bigskip

\noindent {\it Proof of theorem 1.3.} Let $A\in {\cal
W}_2$. Set $A_n= i_{\Lambda_n \Z} \circ \rho _{\Z , \Lambda _n}
(A)$. The  $A_n$ are in ${\cal W}_2$ with finite supports and verify:
$\Vert A_n \Vert _{{\cal W}_2} \leq \Vert A \Vert
_{{\cal W}_2}$. If $m<n$ then we have from  Proposition 2.5:
%---
$$ \Vert A_m - A_n \Vert _{{\cal L}({\cal H}^2 , {\cal H})} \leq
 \Vert \rho _{ \Lambda _n , \Lambda _m } (A_n) - A_n
 \Vert _{{\cal L}({\cal H}^2 , {\cal H})} \leq
 C \sum _{\lambda \in \Z \setminus \Lambda _m \atop  1 \leq j+k \leq 2 }
 \Vert (ad P_{\lambda})^j (ad Q_{\lambda})^k A \Vert _{{\cal L}({\cal
H})}$$
 %-----
 The latter sequence goes to $0$ when $m \rightarrow \infty $ if $A\in
 {\cal W}_2$. Consequently, the sequence $A_n$ converges,
  in ${\cal L}({\cal H}^2 , {\cal H})$, to an element
 $B \in {\cal L}({\cal H}^2 , {\cal H})$.  From
 Theorem 1.4, $B$ is in ${\cal W}_2$ and $A_nf $ strongly converges to
 $Bf$ for all $f\in {\cal H}$.  Let us check that $B=A$.
 To this end, set $f$ and $g$ two elements of ${\cal D}$.
 If $\Lambda _n$ contains the support of $f$ then
 $A_nf = \pi _{\Lambda_n \Z} \  \pi _{\Lambda_n \Z} ^{\star} Af$.
Therefore, if $\Lambda _n$ also contains the support of
$g$:
%----
$$ \big < A_nf , g \big >=\big < \pi _{\Lambda_n \Z} \ \pi
_{\Lambda_n \Z} ^{\star} Af\ , \ \pi _{\Lambda _n \Z} \ \pi _{E_2
\Lambda_n} \psi \big > = \big < \pi _{\Lambda_n \Z} ^{\star} Af\ ,
\  \pi _{E_2 \Lambda_n} \psi \big > = \big < Af , g \big >$$
%------
Since $A_nf$ strongly converges to $Bf$ then $< Af , g> = < Bf ,
g>$ for all $f$ and $g$ in ${\cal D}$. Since ${\cal
D}$ is dense in ${\cal H}$ the equality $B= A$ is indeed true.
As a consequence $A_n$ converges
to $A$ in ${\cal L}({\cal H}^2 , {\cal H})$ and the proof is finished.
 \hfill \carre

\bigskip

Proposition 2.5 also implies the following result.

  \bigskip
  \noindent
  {\bf Corollary 2.6.} {\it For all $A$ and $B$ in ${\cal W}_2$ with
finite supports, one has:
  %-----
  $$ \Vert [A , B] \Vert _{{\cal L}({\cal H}^2 , {\cal H})} \leq C
  \Vert B \Vert _{{\cal W}_2} \ \sum _{\lambda \in \sigma (B) \atop  1
\leq j+k \leq 2 }
 \Vert (ad P_{\lambda})^j (ad Q_{\lambda})^k A \Vert _{{\cal L}({\cal
 H})} \leqno (2.17)$$
 %-----
 where $C$ is not depending on any of the parameters.
}

\bigskip
\noindent {\it Proof.} We make use of the operator $\rho
_{F E}$ for $F = \sigma (A) \cup \sigma (B)$ and $E= F \setminus
\sigma (B)$.  It is known that $\rho _{FE}(A)  $ commutes with $B$ since
its support does not intersect  $\sigma
(B)$. It is then deduced that:
%-----
$$\Vert [ A\ , \ B ]  \Vert_{{\cal L}({\cal H}^2 , {\cal H})} \ =\
\Vert [  A - \rho_{FE}  (A)  \ , \ B ]  \Vert_{{\cal L}({\cal H}^2
, {\cal H})}$$
%----
$$\ \leq \Big [  \Vert B \Vert _{{\cal L} ({\cal H}^2)} +
 \Vert B \Vert _{{\cal L} ({\cal H})} \Big ] \  \Vert A
  - \rho_{FE} (A)
\Vert_{{\cal L}({\cal H}^2 , {\cal H})}$$
%-----
From proposition 2.1,
%----
$$\Vert B \Vert _{{\cal L} ({\cal H}^2)} +
 \Vert B \Vert _{{\cal L} ({\cal H})} \leq C \Vert B \Vert _{{\cal
 W}_2}$$
 %-----
Using Proposition 2.5, we find  a
constant $C>0$, which does not depend on any of the parameters,  such that
(2.17) is verified.

\hfill \carre

\bigskip

\noindent {\bf 3. Perturbation potentials and commutators.}

\bigskip

We have to express the   perturbation potentials
$V_{\lambda}$ and $V_{\lambda \mu}$, satisfying hypotheses
(H1) and (H2) in section 1, as integrals of the Weyl operators, and to
verify precisely that, under our  hypotheses (H1) and (H2), these integrals
are convergent and define  operators in Sobolev spaces.
We shall do the same work  for the commutators of $V_{\lambda \mu}$
with  elements of ${\cal W}_1$, or with Segal operators, or for
iterated commutators.
These norm estimates
will  be used in following sections.
\bigskip
\noindent {\it  Partial  Sobolev spaces.}
\bigskip
  The  Sobolev spaces defined in section 2 are not
  Hilbert spaces. Nevertheless, for any finite subset like
  $\Lambda_n$, the space ${\cal H}^k_{\Lambda_n}$ may be endowed with
    an Hilbert space norm which is equivalent, for each fixed $n$,
    to the norm of section 2. As an example, for $k=1$,
one may set:
  %-----
  $$ \Vert f \Vert _{{\cal H}^1_{\Lambda_n}}^2 = \sum _{\lambda
  \in \Lambda_n \atop j=0, 1} \Vert Q_{\lambda}^{(j)} f \Vert
  _{{\cal H}_{\Lambda_n}}^2$$
  %-----
 For all $n$, these norms and those on section 2 are equivalent but the
constant involved in the inequality depends on $n$.
  \bigskip
 Let us choose
  an orthonormal basis $(\varphi _{\alpha} )_{(\alpha \geq 0)}$ in the
Hilbert space ${\cal H}_{\Lambda_n^c}$.
 We define a map $\Psi _{\alpha} $ from ${\cal H}
  _{\Lambda_n}$ into ${\cal H}$ by $\Psi _{\alpha} (f) = f
  \otimes \varphi _{\alpha}$. The adjoint map
  from ${\cal H}$ to ${\cal
  H}_{\Lambda_n}$ is denoted by $\Psi _{\alpha}^{\star}$. For all $f$ in
${\cal H}$ we have:
  %-----
$$ \Vert f \Vert ^2 = \sum _{\alpha \geq 0} \Vert \Psi
_{\alpha}^{\star} f \Vert _{{\cal  H}_{\Lambda_n}}^2$$
%-----
Then, we define the space ${\cal H}^k(\Lambda_n)$ as the set of all  $f$
with a  finite below norm:
%-----
$$ \Vert f \Vert _{{\cal H}^k(\Lambda_n)}^2 = \sum_{\alpha \geq 0}
\Vert  \Psi _{\alpha}^{\star} f \Vert _{{\cal
H}^k_{\Lambda_n}}^2 \leqno (3.1)$$
%------
Thus, ${\cal H}^k \subset {\cal H}^k(\Lambda_n) \subset {\cal H}$ if
$k\geq 0$. When $k=1$, an element $f$ of ${\cal H}$ is in
${\cal H}^1$ if it belongs to ${\cal
H}^1(\Lambda_n)$ and if, for all $\lambda \in \Lambda _n^c$, one has
$Q_{\lambda}^{(j)}f \in {\cal H}$, the sequence of these norms being
bounded. This property may be used only for fixed $n$.

\bigskip
\noindent {\it  Partial Sobolev spaces with negative order.}

 Set
${\cal H}^{-k} (\Lambda _n)$ the anti-dual set of
  ${\cal H}^k (\Lambda _n)$ ($k=1 , 2$).
Thus:
  %-------
   $${\cal H}^2 (\Lambda _n) \subset {\cal H}^1 (\Lambda _n) \subset
   {\cal H} \subset {\cal H}^{-1} (\Lambda _n) \subset
   {\cal H}^{-2} (\Lambda _n)$$
   %-----
  If an operator $\Phi \in {\cal L} ({\cal H}^1 _{\Lambda _n}  ,
  {\cal H})$ verifies $\big < \Phi f , g \big > = \big < f , \Phi g
  \big >$ for all $f$ and $g$ in ${\cal H}^1 _{\Lambda _n}$,
   where $\big < .\ , \ . \big >$ is the scalar product in ${\cal
   H}$,  then,  for all $f\in {\cal H}$,  the map $g
   \rightarrow  \big < f , \Phi g \big >$ is an element of
  $ {\cal H}^{-1} (\Lambda _n)$ denoted here by $\Phi f$. Thus,
  the operator $Q_{\lambda}$ is bounded from ${\cal H}^k (\Lambda_n)$
  into ${\cal H}^{k-1} (\Lambda_n)$ ($-1 \leq k\leq 2$ , $\lambda \in
  \Lambda_n$). We shall check that similar considerations are also valid
for the operators
  $i [ P_{\lambda} ,
   V_{\Lambda _n} ]$. The commutator of these two types of operators is
   in  ${\cal L} ({\cal H}^1 (\Lambda _n) ,{\cal H}^{-1} (\Lambda
   _n))$.

\bigskip

\noindent {\it  Perturbation potentials and Weyl operators.}
\medskip
If $\xi$ is real sequence in $\ell^2(\Z)$ with a finite support then the
Segal operator  $\Pi (\xi , 0)$ defined in (2.1)
is also written as $\sum \xi_{\lambda} Q_{\lambda}$. Since the
hypotheses on the  perturbation potentials involve only
the derivatives of order 2 and 3,  the following function shall be
implied in the sequel:
%-----
$$x \rightarrow F(x) = e^{ix}-1 -ix = i^2 x^2 \int_0^1 (1-\theta)
e^{i\theta x} d\theta \leqno (3.2)$$
%------
Set $V_{\lambda_1 , \lambda_2}$ ($\lambda_1 \not= \lambda_2$)
the non  bounded operator in ${ \cal H}_{ \{ \lambda _1
\lambda_2 \} }$ being, under the identification of this space with
$L^2(\R^2)$, the multiplication by a function $v_{\lambda_1
\lambda_2}$. If the latter  satisfes the hypothesis (H1), one
has:
%----
$$ V_{\lambda_1 \lambda_2} =  v_{\lambda_1 \lambda_2}(0) \ I + \sum
_{1 \leq j \leq 2} ( \partial _{\lambda _j} v_{\lambda_1
\lambda_2}) (0) \ Q_{\lambda_j} \ +  (2 \pi)^{-2} \int _{\R^2}
\widehat { v_{\lambda_1 \lambda_2}} (\xi) \  F\big ( \xi_{\lambda_1}
Q_{\lambda_1 } + \xi_{\lambda_2} Q_{\lambda_2 }\big )\ d\xi \leqno
(3.3)  $$
%-------
Under the hypothesis (H1) the integral is convergent and it
defines a bounded operator from ${\cal H}^2$ in ${\cal H}$.

\bigskip
\noindent {\it Commutators.}
%--------
\medskip
 In order to study the commutators of  $
V_{\lambda_1 \lambda_2} $ with other operators, we shall use the following
relations,  valid for any operators $X$ and $A$ in a Banach space,
and for the function $F$ in (3.2):
%-----
$$ [e^{iX} , A ] = i \int_0^1  e^{i\theta X} \ [X , A] e^{i(1 -
\theta) X}\ d\theta \leqno (3.4)$$
%-------
$$ [F(X), A ] = i [X , A] \ (e^{iX} - I) \ + \ i^2 \int _0^1 (1-
\theta) e^{i\theta X} \ \big [\ X ,  [X , A ]\ \big ] e^{i (1 -
\theta)  X} d\theta.\leqno (3.5)$$
%------
\bigskip
Equality (3.5) is first applied with $X =
\xi_{\lambda_1} Q_{\lambda_1 } + \xi_{\lambda_2} Q_{\lambda_2 }$
and $A= P_{\lambda_j}$ ($j= 1 , 2$). Using equality
(3.3) for $ V_{\lambda_1 \lambda_2}$ we obtain:
%---------
$$ [P_{\lambda_j}  ,  V_{\lambda_1 \lambda_2} ] = -i \big (
\partial _{\lambda _j} v_{\lambda_1 \lambda_2} \big ) \big (0 \big
) I \  + \ \sum _{1 \leq k \leq 2}  A_{\lambda _1 \lambda_2}^{jk}
Q_{\lambda_k}$$
%------
$$  A_{\lambda _1 \lambda_2}^{jk} = (2 \pi)^{-2} \int _{\R^2
\times [0, 1]} \widehat { v_{\lambda_1 \lambda_2}} (\xi)\
\xi_{\lambda_j} \xi_{\lambda_k}  \ e^{i \theta (\xi_{\lambda_1}
Q_{\lambda_1 } + \xi_{\lambda_2} Q_{\lambda_2 })} d\xi d\theta$$
%----
Under the assumption (H1), this integral converges and defines
an operator $ A_{\lambda _1 \lambda_2}^{jk}$ in ${\cal
L}({\cal H})$, with a norm being  ${\cal O} ( e^{-\gamma_0 |
\lambda_1 - \lambda_2 |} )$. Each one site operator is similarly treated.
Note that the integrals are then integrals on
$\R$. We deduce the following proposition concerning  the potential
$V_{\Lambda _n}$ defined in (1.1) and (1.9):

\bigskip
 \noindent {\bf Proposition 3.1.} {\it Under the hypotheses (H1) and
 (H2),  one may write:
%---
$$ [ P_{\lambda}, V_{\Lambda _n} ] = -i a_{\lambda}^{(n)} + \sum
_{\mu \in \Lambda _n} W_{\lambda \mu}^{(n)} \  Q_{\mu} \leqno (3.6)$$
%----
where $a_{\lambda}^{(n)} $ is a real constant number,  and where
$W_{\lambda \mu}^{(n)}$ is a bounded operator in ${\cal H}$. Moreover,
there exists $C_1>0$ independent of $\lambda$, $\mu$  and $n$,
such that:
%----
$$ | a_{\lambda}^{(n)}| \leq C_1,  \hskip 1cm \Vert  W_{\lambda
\mu}^{(n)} \Vert_{{\cal L}({\cal H})} \  \leq \ C_1 e^{-\gamma_0 |
\lambda - \mu |}.  \leqno (3.7)$$
%------
}

\bigskip

We can also apply the commutation formula (3.5), still setting
$X=\xi_{\lambda_1} Q_{\lambda_1 } + \xi_{\lambda_2}
Q_{\lambda_2 }$, but with $A \in {\cal W}_2$. Inserting the expression
(3.3) for $V_{\lambda_1 \lambda_2}$ and using hypotheses (H1), we obtain
the following  proposition.

\bigskip

\noindent {\bf Proposition 3.2.} {\it For all $A$ in ${\cal
W}_2$, for all $\lambda$ and $\mu$ in $\Z$, the commutator $ [A
, V_{\lambda \mu} ]$ is in ${\cal L}( {\cal H}^1 , {\cal H})$.
There is $C>0$, independent of all the parameters, such that:
%----
$$ \Vert [A , V_{\lambda \mu} ]  \Vert_{{\cal L} ({\cal H}^1 ,
{\cal H})}\  \leq C e^{-\gamma_0|\lambda - \mu |} \sum _{1 \leq j
+ k \leq 2 } \Vert (ad \  Q_{\lambda })^j \ (ad \  Q_{\mu })^k A ]
 \Vert_{{\cal L} ({\cal H})}  \leqno (3.8)$$
%-------
}

\bigskip \noindent {\it  Double commutators.}
\medskip
If $A$, $B$ and $X$ are three operators such that $[X , B]$ is the
identity operator up to a multiplicative factor, and if $F$ is the function
given by  (3.2),  then
it is deduced from  (3.4) and  (3.5) that:
%-----
$$ \big [ \ [ F(X) , B ], A \big ] = i^2 [X , B] \int_0^1 e^{i
\theta X} \ [X , A] \ e^{i (1-\theta)X} d\theta \leqno (3.9) $$
%----------
This formula is applied with  $X=\xi_{\lambda_1}
Q_{\lambda_1 } + \xi_{\lambda_2} Q_{\lambda_2 }$, $B=
P_{\lambda_j}$ ($j=1 , 2$) and $A \in {\cal L} ({\cal H})$ (in particular
$A \in {\cal W}_1$). Inserting the expression
(3.3) for $V_{\lambda_1 \lambda_2}$ and using the hypotheses
(H1), on gets:
%-------
$$ \big [ \ [ V_{\lambda_1 \lambda_2} , P_{\lambda_j} ] , A \big ]
\ =  \ \sum _{1 \leq k \leq 2}
 S _{\lambda _1 \lambda_2}^{jk} \big ( [A ,Q_{\lambda_k}] \big ) \leqno
 (3.10)$$
 %----
 where we set, for all $\Phi$ in ${\cal L} ({\cal H}^2(\Lambda_n) ,
 {\cal H}^{-2}(\Lambda_n))$
 %------
 $$ S _{\lambda _1 \lambda_2}^{jk} \big ( \Phi  \big ) =
  (2 \pi)^{-2} \int _{\R^2
\times [0, 1]} \widehat { v_{\lambda_1 \lambda_2}} (\xi)\
\xi_{\lambda_j} \xi_{\lambda_k}  \ e^{i \theta X(\xi)}\ \circ \Phi
\circ e^{i (1-\theta)X(\xi)} d\xi d\theta \leqno (3.11)$$
%-----------
with the notation $X(\xi) = \xi_{\lambda_1} Q_{\lambda_1 } +
\xi_{\lambda_2} Q_{\lambda_2 }$.

\bigskip
Next we shall deduce the following proposition.
\bigskip
 \noindent {\bf Proposition 3.3.} {\it
 For all $\lambda$ and  $\mu$ in $\Lambda_n$ ($n\geq 1$),
there exists a linear continuous map $ K_{\lambda \mu}$ from ${\cal L}
({\cal H}^2(\Lambda_n),$
 ${\cal H}^{-2}(\Lambda_n))$ into itself, leaving invariant the
subspaces ${\cal L} ({\cal H}^1(\Lambda_n) , {\cal H}^{-1})
  (\Lambda_n)$ and ${\cal L} ({\cal H})$,
   such that, for all $A$ in ${\cal L} ({\cal H})$,
we have:
%----
$$\big [ A ,  [ P_{\lambda}, V_{\Lambda _n} ]\ \big ] = \sum _{\mu
\in \Lambda_n} K_{\lambda \mu} \big ( [A , Q_{\mu} ] \big ) \leqno
(3.12)$$
%----
Moreover,   when restricted to ${\cal L}
({\cal H})$, $ K_{\lambda \mu}$  is in ${\cal L}( {\cal L} ({\cal
H}))$,
and there exists $C_0>0$, independent of $n$, $\lambda$ and $\mu$,  such that
%---
$$ \Vert K_{\lambda \mu} \Vert _{ {\cal L} ( {\cal L} ({\cal H}))}
\ \leq C_0 e^{-\gamma_0|\lambda - \mu|}.  \leqno (3.13)$$
%----
}

\bigskip \noindent {\it Proof.}
Under our hypotheses, the operator   $\Phi \rightarrow  S
_{\lambda _1 \lambda_2}^{jk} \big ( \Phi  \big )$ maps ${\cal L}
({\cal H}^2(\Lambda_n) , {\cal H}^{-2}(\Lambda_n))$ into itself. It also
maps ${\cal
 L}({\cal H})$ into itself, with a norm  $\leq C_0
 e^{-\gamma_0|\lambda - \mu|}$.
For one site potentials $V_{\lambda}$, we define similar
 operators $S_{\lambda}$ such that
$\big [ \ [ V_{\lambda }, P_{\lambda} ] , A \big]  = S_{\lambda} \big ( [
A ,
Q_{\lambda}] \big ) $, for all $A\in {\cal W}_1$.
We then set, for all $\lambda$ and
$\mu$ in $\Lambda_n$ such that $\lambda \not= \mu$:
%----
$$K_{\lambda \mu }^{(n)}(\Phi)= \left \{ \matrix { S_{\lambda
\mu}^{12} (\Phi)+ S_{\mu \lambda}^{21}(\Phi) &{\rm if}& |\lambda -
\mu| \geq 2\cr -b\Phi + S_{\lambda \mu}^{12}(\Phi) + S_{\mu
\lambda}^{21}(\Phi) &{\rm if}& |\lambda - \mu| = 1 \cr } \right
.$$
%----
and if $\lambda = \mu$,
%------
 $$ K_{\lambda \lambda }^{(n)} (\Phi) =  a \Phi +
S_{\lambda}(\Phi) + \sum_{\mu \in \Lambda _n \atop \mu \not=
\lambda} ( S_{\lambda \mu}^{11}(\Phi) + T_{\mu \lambda}^{22}(\Phi)
)$$
%----
The equality (3.12) and the estimates (3.13) follows.

\hfill   \carre

\bigskip

\bigskip
A consequence  of proposition 3.2, (that shall be used in the sequel), is
that the left product by the matrix $\Vert K_{\lambda
\mu} (t) \Vert $ leaves invariant the set of matrices with exponential
decay. In particular, it is precisely the function $S_{\gamma}$ implied in
the next
 proposition which will determine the propagation speed in section 8.

\bigskip

\noindent {\bf Proposition 3.4.} {\it  Under the hypotheses (H1)
and (H2) and for all $\gamma $ in $]0, \gamma _0[$ (or in $]0,
\infty[$ in the case of interaction with nearest neighbors),
there exists $S_{\gamma} >0$ such that, for all $n$, for all
$\lambda$ and $\nu$ in $\Lambda_n$:
%----
$$ \sum _{\mu \in \Lambda _n} \Vert K _{\lambda \mu}  \Vert
_{{\cal L}({\cal L}({\cal H}))} e^{-\gamma |\mu - \nu|} \leq
S_{\gamma} e^{- \gamma |\lambda - \nu|} \hskip 1cm   \sum _{\mu
\in \Lambda _n} \Vert W _{\lambda \mu}  \Vert _{{\cal L}({\cal
H})} e^{-\gamma |\mu - \nu|} \leq S_{\gamma} e^{- \gamma |\lambda
- \nu|} $$
%----
where the $K _{\lambda \mu}$ are the operators constructed in
Proposition 3.3 and where the  $W _{\lambda \mu}$ are those of Proposition
3.1.

 }

 \bigskip

\bigskip
\noindent {\it Triple commutators.}
\medskip
\medskip
If $X$, $A$, $B$, $C$ are operators such that $[X , B]$ and
$[X , C]$ are equal to the identity operator up to a multiplicative
factor, and if $F$ is the function defined by (3.2), then we deduce from
(3.9) and (3.4) that:
%-----
$$ \Big [ \ \big [ \ [ F(X) , B] , A \big ] , C \Big ] = i^2 [X ,
B] \int_0^1 e^{i \theta X} \ \big [\ [X , A] , C \big ]  \ e^{i
(1-\theta)X} d\theta  \ +\ i^3 [X , B] \ [X , C] \int_0^1 e^{i
\theta X} [X , A]  \ e^{i (1-\theta)X} d\theta $$
%-------
We shall apply this formula with  $X=\xi_{\lambda_1}
Q_{\lambda_1 } + \xi_{\lambda_2} Q_{\lambda_2 }$, $B=
P_{\lambda_j}$,($1 \leq j  \leq 2$),  $A \in {\cal W}_2$ and $C$
being a Segal operator. Inserting the expression of $V_{\lambda_1
\lambda_2}$ given in
(3.3) and using the hypothesis
(H1),  we obtain:
%------
$$  \Big [ \ \big [ \ [ V_{\lambda_1 \lambda_2}  , P_{\lambda_j}]
, A \big ] ,  C \Big ] = \sum _{1\leq k \leq 2}
 S _{\lambda _1 \lambda_2}^{jk } \Big (\big [\  [A ,Q_{\lambda_k}],
 C  \Big ) +  \  T_{\lambda _1 \lambda_2}^{jk } \Big (  [A
 ,Q_{\lambda_k}]  , C \Big)$$
 %------
 where  $ S _{\lambda _1 \lambda_2}^{jk }(\Phi)$ is the operator
defined in (3.11) and $T_{\lambda _1 \lambda_2}^{jk } (\Phi , C)$
is defined by:
 %-------
 $$ T_{\lambda _1 \lambda_2}^{jk } (\Phi , C) =
(2 \pi)^{-2} \int _{\R^2 \times [0, 1]} \widehat { v_{\lambda_1
\lambda_2}} (\xi)\ \xi_{\lambda_j}  \xi_{\lambda_k} [ X(\xi) , C]
\ e^{i \theta X(\xi)}\ \circ \Phi \circ e^{i (1-\theta)X(\xi)}
d\xi d\theta$$
%-----
If $C$ is Segal operator  (linear combination of
$P_{\lambda}$ and $Q_{\lambda}$) then $[ X(\xi) , C]$  is a constant
and the above integral  converges using  the hypothesis (H1).
It is at this point that the  hypothesis:  "$|\xi|^3 \widehat {
v_{\lambda \mu}} (\xi)$ belongs to $L^1(\R^ 2)$" is involved. We proceed
similarly for all one site operators  $V_{\lambda}$. Summing up as in
Proposition 3.3, one obtains the next result:

\bigskip

\noindent {\bf Proposition 3.5.}  {\it For all $\lambda$ and
$\mu$ in $\Lambda_n$ ($n\geq 1$), for all Segal operator
 $\Psi$, there exists a map $\Phi \rightarrow
R_{\lambda \mu} (\Phi , \Psi)$ from ${\cal L}( {\cal H}^1
(\Lambda_n) , {\cal H}^{-1} (\Lambda_n))$ into itself such that, for all
$A\in {\cal L}( {\cal H})$ supported in
$\Lambda_n$, we have:
%-----
$$ \Big [ \ \big [ A , [ P_{\lambda} , V_{\Lambda_n} ]  \big ] \
,\ \Psi  \ \Big ] \ =\ \sum _{\mu \in \Lambda_n} K_{\lambda , \mu}
\Big ( \big [  [A , Q_{\mu} ] , \Psi \big ] \Big ) \ +\ R_{\lambda
, \mu} \big (  [A , Q_{\mu} ] , \Psi \big) \leqno (3.)$$
%--------
where $\Phi \rightarrow  K_{\lambda , \mu} (\Phi)$ is the map of
Proposition 3.3. If $\Phi $ is in ${\cal
L}({\cal H})$ then $R_{\lambda \mu } (\Phi , \Psi )$ is in
${\cal L}({\cal H})$. One has $R_{\lambda \mu } ( \Phi , Q_{\rho} ) =
0$ for all $\rho$. One also see that $R_{\lambda \mu } (\Phi ,
P_{\rho}  ) = 0$ excepted when the set $\{ \lambda , \mu , \rho \}$
has only two distinct elements ($\lambda = \mu$ or $\lambda=
\rho$ or $\mu= \rho$). In that case, one gets:
%----
$$ \matrix { \Vert R_{\lambda \mu } (\Phi  ,  P_{\rho} )\Vert
_{{\cal L}( {\cal H})} \leq C_0 e^{-\gamma _0 |\lambda - \mu|}
\Vert \Phi \Vert _{{\cal L}( {\cal H})}  & {\rm if}& \lambda \not=
\mu \cr \Vert  R_{\lambda \mu } (\Phi ,  P_{\rho} )\Vert _{{\cal
L}( {\cal H})} \leq C_0 e^{-\gamma _0 |\lambda - \rho|} \Vert \Phi
\Vert _{{\cal L}( {\cal H})}  & {\rm if}& \lambda = \mu \cr } $$
%---------
}

\bigskip

\noindent {\bf 4.  Evolution of  the  position and impulsion  operators.}

\bigskip
Using the Fock space notations, the Hamiltonian
  $H_{\Lambda_n}$ in (1.1) is written as:
%------
$$H_{\Lambda_n} =  \sum _{\lambda \in \Lambda _n} \Big [
P_{\lambda}^2 +  { a \over 2} Q_{\lambda}^2 \Big ] \ - \ b \sum
_{\lambda = -n}^{n-1} Q_{\lambda} Q_{\lambda +1} +
V_{\Lambda_n}^{pert} \leqno (4.1)$$
%------
where the operator $V_{\Lambda_n}^{pert}$ is expressed as the sum (1.9).
The terms in the sum  verify the hypotheses (H1) and
(H2) and let us recall that these two hypotheses are analyzed in
section 3. Let us first start by giving the domain of self-adjointness of
$H_{\Lambda_n}$.
%----
\bigskip

\noindent {\bf Proposition 4.1. } {\it In the Hilbert space
${\cal H} _{\Lambda _n}$, the operator $H_{\Lambda_n}$ is self-adjoint
with the
domain ${\cal H}^2 _{\Lambda _n}$. The operator
$ e^{itH_{\Lambda _n} }$ is bounded in ${\cal H}^k _{\Lambda
_n}$ ($k=0, 1 , 2$). The operator $ e^{itH_{\Lambda _n} } \otimes
I_{\Lambda_n^c}$ is bounded in ${\cal H}^k (\Lambda _n)$
defined in section 3 ($-2 \leq k \leq 2$).
%----
}

  \bigskip

\noindent {\it Proof.} We know that ${\cal H} _{\Lambda
_n}$ is naturally identified to $L^2(\R^{\Lambda_n}) =
L^2(\R^{2n+1})$ in such a way that the operators $P_{\lambda}$
and $Q_{\lambda}$ become:
%---
$$ P_{\lambda} = {1 \over i} {\partial \over \partial x_{\lambda}}
\hskip 1cm Q_{\lambda} = x_{\lambda}$$
%----
The spaces ${\cal H}^k _{\Lambda _n}$ are then identified to the usual
spaces
    $B^k$ of the theory of  globally elliptic operators
   (cf Helffer [HE]). When
   $V_{\Lambda_n}^{pert}=0$,  the operator $H_{\Lambda_n}$ is a
Schr\"odinger operator, where the potential is a definite positive
quadratic form (if $a>2b>0$). In this case, it is well-known that
$H_{\Lambda_n}$ is self-adjoint with domain $B^2 = {\cal H}^2 _{\Lambda
   _n}$. Let us show that the addition of  $V_{\Lambda_n}^{pert}$
   does not modify this result. With the preceding identification and
under our hypotheses, $V_{\lambda}$ and $V_{\lambda \mu}$ are
multiplications by the functions  $v_{\lambda}$ and $v_{\lambda
   \mu}$ with second-order derivatives going to $0$ at infinity.
(These functions are Fourier transforms of functions being in $L^1(\R)$
or in $L^1(\R^2)$.) Consequently, these functions $v_{\lambda}
(x_{\lambda} ) /| x_{\lambda}|^2$ and $v_{\lambda \mu }
(x_{\lambda}, x_{\mu}) /| x_{\lambda}|^2 + | x_{\mu}|^2$ goes
to $0$ at infinity. The above Proposition thus follows from
Kato-Rellich's theorem. As a consequence, the operator $ e^{itH_{\Lambda
_n} }$ is a well-defined bounded operator in ${\cal H}$ and in the domain
of $H_{\Lambda_n}$, that is to say in  ${\cal H}^2
_{\Lambda _n}$. By interpolation it is also bounded in
${\cal H}^1 _{\Lambda _n}$.
The latter statement comes from (3.1) if $0 \leq k \leq 2$ and is deduced
by duality if $k\leq 0$.

\hfill \carre

\bigskip

Consequently, if  $A$ belongs to ${\cal L}({\cal H}^{k}(\Lambda_n) , {\cal
H}^{k'} (\Lambda_n) )$ then the operator
%-----
$$ \alpha^{(t)}_{\Lambda_n} (A) = ( e^{itH_{\Lambda _n} }\otimes
I) \circ  A\ \circ ( e^{-itH_{\Lambda _n} }\otimes I)  \leqno
(4.2)$$
%---
in the same spaces. In particular, the operator
$ \alpha^{(t)}_{\Lambda_n} (Q_{\lambda}^{(j)})$ ($\lambda \in
\Lambda_n)$  belongs to ${\cal L}({\cal H}^{1}(\Lambda_n) , {\cal
H})$.

\bigskip

\noindent {\bf Proposition 4.2. }  {\it For all $\lambda $ and
$\mu$ in $\Lambda_n$, there exists  $C^1$ maps $t \rightarrow
A_{\lambda \mu}^{(n)} (t)$, $t \rightarrow B_{\lambda \mu}^{(n)}
(t)$, and  $t \rightarrow R_{\lambda }^{(n)} (t)$   from
$\R$  into ${\cal L} ({\cal H})$ such that (omitting the superscript
$n$ in the expressions):
 %----
 $$ \alpha^{(t)}_{\Lambda_n} \big ( Q_{\lambda} \big ) =
 \sum _{\mu \in \Lambda _n} \Big [ A_{\lambda \mu} (t)
 Q_{\mu} +  B_{\lambda \mu} (t) P_{\mu} \Big ] + R_{\lambda }
 (t)\leqno (4.3)$$
 %----
$$ \alpha^{(t)}_{\Lambda_n} \big (  P_{\lambda} \big )  =
 \sum _{\mu \in \Lambda _n} \Big [ A'_{\lambda \mu} (t)
 Q_{\mu} +  B'_{\lambda \mu} (t) P_{\mu} \Big ] + R'_{\lambda }
 (t) \leqno (4.4)$$
 %----
Moreover, for all $\gamma $ in $]0, \gamma_0[$, for all $M>
\sqrt {S_{\gamma}}$, (where $S_{\gamma}$ is the constant number appearing
in
Proposition 3.3), there exists $C>0$ such that:
 %--
$$ \Vert  A_{\lambda \mu} (t)  \Vert + \Vert  B_{\lambda \mu} (t)
\Vert + \Vert  A'_{\lambda \mu} (t)  \Vert +  \Vert B'_{\lambda
\mu} (t)  \Vert  \leq C e^{M | t|} e^{- \gamma | \lambda - \mu |}
 \leqno (4.5)$$
%-------
$$ \Vert  R_{\lambda } (t)  \Vert + \Vert  R'_{\lambda } (t) \Vert
\leq C e^{M | t|}  \leqno (4.6) $$
%-------
 }

\bigskip
\noindent {\it First step.} We shall study the differential system
satisfied by:
%-----
$$ Q_{\lambda}(t) = \alpha^{(t)}_{\Lambda_n} \big (
Q_{\lambda}\big )   \hskip 1cm P_{\lambda}(t) =
\alpha^{(t)}_{\Lambda_n} \big (  P_{\lambda} \big )$$
%----
One observes that  $t \rightarrow  Q_{\lambda}(t)$
and $t \rightarrow  P_{\lambda}(t)$  are $C^1$ functions
from $\R$ into ${\cal L} ({\cal H}^1 (\Lambda_n) ,
{\cal H})$ verifying:
%----
$$  Q'_{\lambda}(t)= P_{\lambda}(t) \hskip 1cm  P'_{\lambda}(t)=
- i \alpha^{(t)}_{\Lambda_n} \big ( [ P_{\lambda} , V_{\Lambda_n}]
\big )$$
%-------
With the operators $W_{\lambda \mu}^{(n)}$
and the constant in $a_{\lambda}^{(n)}$ of Proposition 3.1, it follows
that:
%----
$$ P'_{\lambda}(t)= - a_{\lambda}^{(n)}  \ - i  \  \sum _{\mu \in
\Lambda_n}   \alpha^{(t)}_{\Lambda_n} \big (  W_{\lambda
\mu}Q_{\mu} \big ) $$
 %----
We define an operator ${\cal L} ({\cal H})$ by setting:
%----
$$ \widetilde W _{\lambda \mu} (t)  =
 \alpha^{(t)}_{\Lambda_n} \big (   W_{\lambda \mu}^{(n)}
 \big )  \leqno (4.7)$$
%----
With these notations, the preceding system is written as:
%----
$$ Q'_{\lambda}(t)= P_{\lambda}(t) \hskip 1cm P'_{\lambda}(t)= -
a_{\lambda}^{(n)}  \ -i  \  \sum _{\mu \in \Lambda_n}
 \widetilde W
_{\lambda \mu} (t) \circ  Q_{\mu}(t) \leqno (4.8) $$
 %----
To conclude,  $t \rightarrow
( Q_{\lambda}(t) ,  P_{\lambda}(t) )$ is the unique
 $C^1$ map from $\R$
into ${\cal L} ({\cal H}^1 (\Lambda_n) , {\cal H})$
solution to (4.8) and  satisfying $Q_{\lambda}(0) = Q_{\lambda}$ and
 $P_{\lambda}(0) = P_{\lambda}$.

 \bigskip

\noindent {\it Second step.}  We shall now construct matrices $A_{\lambda
\mu} (t),\dots$ such that the right hand-side of (4.3) is also solution to
the same system (4.8) and satisfies the same initial data.  First, we can
find an operator-valued matrix $(A_{\lambda
\mu} ^0 (t) , A_{\lambda  \mu} ^1 (t)) $ in ${\cal L} ({\cal H})$ solution
to:
%------
$$ {d\over dt} A_{\lambda  \mu} ^0 (t) =   A_{\lambda  \mu} ^1 (t)
\hskip 1cm  {d\over dt} A_{\lambda  \mu} ^1 (t) = -i \sum _{\nu \in
\Lambda _n} \widetilde W_{\lambda \nu} (t)  A_{\nu  \mu} ^0 (t)
\leqno (4.9)$$
%----
$$  A_{\lambda  \mu} ^0 (0) = \delta _{\lambda \mu } I \hskip 1cm
 A_{\lambda  \mu} ^1 (0) = 0$$
 %-----
 Indeed, from Propositions  3.1 and 3.4 one see that the hypotheses in
Proposition B.1 (Appendix B)
 are satisfied for all $\gamma \in ]0, \gamma _0[$.  Then, there
 exists  a solution of (4.9) satisfying the above initial condition,
 and also, if $M > \sqrt { S_{\gamma}}$:
 %-----
 $$ \Vert A_{\lambda  \mu} ^j (t)\Vert_{{\cal L} ({\cal H})} \leq
 C(M, \gamma) e^{M| t|}  e^{-\gamma |\lambda - \mu|} \leqno (4.10)$$
 %------
 An operator-valued matrix  $(B_{\lambda
\mu} ^0 (t) , B_{\lambda  \mu} ^1 (t)) $ solution to the same system (4.9)
verifying the same estimates (4.10) is analogously constructed,
satisfying the following initial conditions:
%----
$$  B_{\lambda  \mu} ^0 (0) =0 \hskip 1cm
 A_{\lambda  \mu} ^1 (0) =  \delta _{\lambda \mu } I $$
 %-----
From remark 2 in the appendix B, one may find
operators $(R_{\lambda}^0 (t) , R_{\lambda}^1 (t) )$ of
${\cal  L}( {\cal H})$ solutions to
%----
$$ {d \over dt} R_{\lambda}^0 (t) = R_{\lambda}^1(t) \hskip 1cm
 {d \over dt} R_{\lambda}^1 (t) = -i  \sum _{\nu \in \Lambda_n }
 \widetilde W _{\lambda \nu} (t) R_{\nu}^0 (t) + i a_{\lambda}^{(n)} $$
 %----
 $$ R_{\lambda}^0 (0) = R_{\lambda}^1 (0) = 0$$
 %------
 $$ \Vert R_{\lambda}^j (t) \Vert_{{\cal L} ({\cal H})} \leq
 C(M, \gamma) e^{M| t|}  \sum _{\mu \in \Lambda _n}
 e^{-\gamma |\lambda - \mu|} |a_{\mu}  | \hskip 1cm j=0, 1$$
%-------
We define the operators of ${\cal L} ({\cal H}^1
(\Lambda_n), {\cal H} )$ by
%----
$$ \widetilde Q_{\lambda}^j (t) = \sum _{\mu \in \Lambda_n}
\Big [ A_{\lambda \mu}^j (t) Q_{\mu} + B_{\lambda \mu}^j (t) P_{\mu}]
+ R_{\lambda _j} (t) \hskip 1cm j=1, 2$$
%------
These functions verify the same system (4.8) as the functions
$Q_{\lambda}^j  (t)$,  together with the same initial conditions
$\widetilde Q_{\lambda}^0  (t) = Q_{\lambda}$,
 $\widetilde Q_{\lambda}^1  (t) = P_{\lambda}$. Uniqueness shows
 $\widetilde Q_{\lambda}^0  (t) = Q_{\lambda} (t)$ and
 $\widetilde Q_{\lambda}^1  (t) = P_{\lambda} (t)$,
thus the equalities (4.3) and (4.4) are true and the matrices estimates
(4.5)
(4.6) are valid.

\hfill   \carre

\bigskip

%--------------EXEMPLE-------------------------

\bigskip
\noindent {\bf Example: The cyclic quadratic case.}

\bigskip
In the case of a positive definite quadratic form potential (without
  perturbation potentials), it is well-known that the equalities (4.3)
and (4.4) are valid with
$R_{\lambda}(t) =0$ and the operators $A_{\lambda \mu} (t)$ and
$B_{\lambda \mu} (t)$  being real numbers. The following classical
proposition may sum up this situation:

\bigskip
\noindent {\bf Proposition 4.3.} {\it In the case where the potentials
$V_{\lambda}$ and $V_{\lambda \mu} $ (perturbation potentials)
are vanishing, the operators $ \alpha^{(t)}_{\Lambda_n} \big (
Q_{\lambda} \big )$ and $ \alpha^{(t)}_{\Lambda_n} \big (
P_{\lambda} \big )$ satisfy equalities (4.2) and (4.3)
where $R_{\lambda}^{(n)} (t) =0$ and the $A_{\lambda \mu}^{(n)}
(t)$ and $B_{\lambda \mu}^{(n)} (t)$ are real numbers. The matrices
$A^{(n)} (t)$ and $B^{(n)} (t)$ are related to the matrix
 $W_n$  of the quadratic form $ V_{\Lambda_n}^{quad}$
 in the canonical basis by the equality :
%----
$$ A^{(n)} (t)=  \cos \big (t   \sqrt {W_n}\big ) \hskip 1cm
 B^{(n)} (t)= -{ \sin \big (t   \sqrt {W_n}\big ) \over  \sqrt
 {W_n}}$$
 %----
}

\bigskip
One may estimate the matricial elements $A_{\lambda \mu} (t)$ and
 $A_{\lambda \mu} (t)$ using Proposition 4.2. However, in some cases, the
inequalities of  Proposition 4.2 together with the Lieb-Robinson
inequalities may be strongly improved and explicitly written down. This is
precisely the case if the perturbation potential vanishes,  and if the
quadratic potential takes the following form (with an interaction between
the two ends of the linear chain):
%---
$$  V_{\Lambda_n}^{cycl} (x)=  { a \over 2} |x|^2 - b  \sum
_{\lambda = -n}^{n-1} x_{\lambda }  x_{\lambda+1} - b  x_n
x_{-n}$$
%----
In that case, we can make the estimates of proposition
4.2 more precise if the distance d$(\lambda ,
\mu)=|\lambda -\mu|$ is replaced by the cyclic distance on $\Lambda_n$,
$d_n (\lambda , \mu) = d( \lambda - \mu , (2n+1)\Z )$.
\medskip
These improved estimates follow on from [NRSS] in the cyclic quadratic
case. Let us give here a simplified proof of a perhaps less precise  type
of estimates.

\medskip
In the cyclic quadratic case,
the analysis of chains of oscillators involves the dispersion relations
$\omega (\theta) = \sqrt { a - 2b \cos
\theta}$ (cf Cohen-Tannoudji [CT]). It is the natural to give a
corresponding complex expression
by setting
%----
$$ \Omega (z) = \sqrt { a  -  b(z +  z^{-1})} \leqno (4.11)$$
%----
This function is analytic in ${\bf C}\backslash \{]- \infty
, z_1]\cup[z_2 , 0]\}$ where $z_1$ and $z_2$ are the roots of
$bz^2 - az + b=0$. Note however that, the function $|{\rm Im} \Omega (z)|$
is well defined on ${\bf C}\setminus \{ 0\}$. Set, for all
$\gamma >0$
%----
$$ M(\gamma) = \sup _{|z| = e^{\gamma}}|{\rm Im}\  \Omega (z)|
\leqno (4.12)$$
%----
This function is well-defined on ${\bf C} \setminus \{ 0
\}$.

\bigskip \noindent {\bf Proposition 4.4.} {\it Under the above hypotheses
and for all $\gamma >0$ there exists $C(\gamma)>0$,
independent on $n$ such that, the matrices $A^{(n)} (t)$ and
$B^{(n)} (t)$ of Proposition 4.3 satisfy:
%----
$$\big | A_{\lambda \mu}^{(n)} (t)\big | \ +\ \big | B_{\lambda
\mu}^{(n)} (t)\big | + \big |{d \over dt} A_{\lambda \mu}^{(n)}
(t)\big |\ +\ \big |{d\over dt} B_{\lambda \mu}^{(n)} (t)\big |\
\leq C(\gamma) e^{|t|M(\gamma)} e^{-\gamma d_n(\lambda , \mu)}$$
%----
where $M(\gamma)$ is defined in (4.12) and $d_n(\lambda , \mu) =
d( \lambda - \mu , (2n+1)\Z )$. }

\bigskip
\noindent {\it Proof.} The matrix $W_n$ of the quadratic form
$V_{\Lambda_n}^{cycl}$, and therefore all the matrices $ A^{(n)}
(t)$ and $ B^{(n)} (t)$ are functions of the cyclic shift operator $S_n$
defined in $\R^{\Lambda_n}$ by
%----
$$ S_ne_j = \left \{ \matrix { e_{j+1}&{\rm if}&  -n \leq j < n
\cr e_{-n} &{\rm if}& j= n \cr } \right .$$
%----
More precisely, one has $W _n= aI + bS_n + b S_n^{-1}$ and
%-----
$$A^{(n)}(t)=  f(S_n , t) \hskip 1cm B^{(n)}(t) = g(S_n , t)
 \hskip 1cm C^{(n)}(t) = h(S_n, t)$$
 %----
where we set, using the function $\Omega (z)$ defined in (4.11):
 %----------
$$ f(z, t) = \cos (t  \Omega (z)) \hskip 1cm  g(z , t)= {\sin  ( t
\Omega (z)) \over \Omega (z) } \hskip 1cm h(z , t) = - \sin  ( t
\Omega (z) ) \Omega (z) \leqno (4.13)$$
%---
These functions are analytic on ${\bf C} \setminus \{ 0 \}$. The proof
uses the following elementary lemma:

\bigskip
\noindent {\bf Lemma 4.5.} { \it Let $S$ be a unitary operator in an
Hilbert space ${\cal H}$. Set $f(z , t)$ the function defined in
(4.11) and (4.13) where $a> 2 | b |>0$. Then, one may write for all $t\in
\R$:
%---
$$ f(S , t)  = \sum _{k \in \Z} c_k(t) S^k $$
%----
Moreover, one has for all $\gamma >0$, for all $t\in \R$ and for all $k\in
\Z$ ,
%----
$$ | c_k(t) |  \leq   e^{- \gamma |k |} \  {1 \over 2 \pi} \int
_0^{2\pi} |f( e^{\gamma}e^{i\theta}  , t)|d\theta $$
%----
The same result holds for the functions $g$ and $h$
defined in (4.13). }

\bigskip
\noindent {\it End of the proof of Proposition 4.4.} Since
$S_n^{2n+1}=I$, the sum in  Lemma 4.5 is written as a finite sum, and
%-----
$$ A^{(n)}(t)= f( S_n , t ) = \sum _{k= 0}^{2n} a _k (t) S_n^k
\hskip 1cm a _k (t) = \sum _{p \in \Z} c_{k +p(2n+1)} (t)$$
%---
where the $c_j (t)$ are the coefficients of Lemma 4.5. Consequently, if
$-n \leq \lambda  \leq \mu \leq n$ and $\gamma
>0$ one has:
%----
$$ | A_{\lambda \mu}^{(n)} (t) | = \Big | \Big <  f( S_n , t )
e_{\lambda} , e_{\mu} \Big > \Big | = |a _{\mu - \lambda }(t)|
\leq  \sum _{p \in \Z} | c_{\mu - \lambda +p(2n+1)} (t) |$$
%----
$$ \leq \left [ \sum _{p\in \Z} e^{-\gamma |\mu - \lambda
+p(2n+1)|}\right ]\  {1 \over 2 \pi} \int _0^{2\pi} |f(
e^{\gamma}e^{i\theta}  , t)|d\theta $$
%-------
There exists $C_1 (\gamma)$ and $C_2 (\gamma)$ independent of $n$,
such that:
%-------
$$\sum _{p\in \Z} e^{-\gamma |\mu - \lambda +p(2n+1)|} \leq
C_1(\gamma) e^{-\gamma d_n ( \lambda , \mu)} $$
%------
$$ {1 \over 2 \pi} \int _0^{2\pi} |f( e^{\gamma}e^{i\theta}  ,
t)|d\theta \leq  C_2(\gamma ) e^{|t| M(\gamma)}$$
%-----
where $M(\gamma)$ is defined in (4.12).  As a consequence, $ |
A_{\lambda \mu}^{(n)} (t) | \leq C_1(\gamma) C_2(\gamma)
 e^{|t| M(\gamma)}  e^{-\gamma d_n ( \lambda , \mu)}$.
 Similar estimates for the matricial elements
$B_{\lambda \mu}^{(n)} (t)$ together with its derivatives may be obtained.
The proof of Proposition
4.4 follows.

\hfill \carre
\bigskip

%-------------FIN--DE L'-EXEMPLE----------------------

\bigskip
\noindent {\bf 5. Evolution of the commutators.}
\bigskip

From  Proposition 4.1, the
commutators  $[A ,  \alpha ^{(t)} _{\Lambda_n} (Q_{\lambda}) ]$
and $[A ,  \alpha ^{(t)} _{\Lambda_n} (P_{\lambda}) ]$ are defined as
operators taking ${\cal H}^1 (\Lambda _n) $ into
${\cal H}^{-1} (\Lambda _n) $, for all $A$ in
 ${\cal L} ({\cal H})$ supported in $\Lambda _n$,
 and for all $t\in \R$.

\bigskip
\noindent {\bf Proposition 5.1.}  {\it For all $A\in {\cal
W}_1$ supported in $\Lambda _n$ and for all $t\in \R$ the
commutators  $[A ,  \alpha ^{(t)} _{\Lambda_n}
(Q_{\lambda}^{(j)}) ]$ are bounded in ${\cal H}$ ( $\lambda \in
\Lambda_n$, $0 \leq j \leq 1$). For all $\gamma$ in the interval
$]0, \gamma_0[ $
and for all $M > \sqrt {S_{\gamma}}$ there exists $C(M, \gamma ) >0$,
(independent of $n$) such that:
%-----
$$ \Vert [A ,  \alpha ^{(t)} _{\Lambda_n} (Q_{\lambda}^{(j)})
]\Vert _{{\cal L}({\cal H})} \leq C (M , \gamma )  e^{M|t|} \sum
_{\mu \in \Lambda _n \atop 0 \leq k \leq 1}e^{-\gamma d (\lambda ,
\mu)}\ \Vert [A , Q_{\mu }^{(k)}] \Vert _{{\cal L}({\cal H})}
\leqno (5.1)$$
 %-------
}

\bigskip

 \noindent {\it First step.}  Assuming first that
 $A$ is only in ${\cal L}({\cal H})$ we shall study the differential
system satisfied by the functions:
%-------
$$ \Phi_{\lambda }^{(j)} (t) = \big [ A , \alpha ^{(t)}
_{\Lambda_n} (Q_{\lambda}^{(j)} ) \big ] \hskip 1cm 0 \leq j \leq
1 \leqno (5.2)$$
%-------
 The $ \Phi_{\lambda }^j$'s are
$C^1$ maps from $\R$ into ${\cal L} ( {\cal H}^{1}
(\Lambda _n) , {\cal H}^{-1}(\Lambda _n) )$ and verify:
%-----
$${d \over dt}  \Phi _{\lambda }^0 (t) =  \Phi _{\lambda }^1 (t)
\hskip 1cm {d \over dt}  \Phi _{\lambda }^1 (t) = -i \Big [ A ,
\alpha ^{(t)} _{\Lambda_n}  \big ( [ P_{\lambda} , V_{\Lambda_n}
]\big )
 \Big ] = - i  \alpha ^{(t)} _{\Lambda_n}  \Big (  \big [ \alpha _{\Lambda
_n}^{(-t)} (A) , [ P_{\lambda} , V_{\Lambda_n} ] \big ] \Big )$$
%------
Using the operators $K_{\lambda \mu}$ of Proposition
3.3,
%------
$$ \big [ \alpha _{\Lambda _n}^{(-t)} (A) , [ P_{\lambda} ,
V_{\Lambda_n} ] \big ] =  \sum _{\mu \in \Lambda _n}  K_{\lambda
\mu} \big ( [ \alpha _{\Lambda
 _n}^{(-t)} (A) , Q_{\mu}   ] \big ) $$
 %-----
Next set $\widetilde K _{\lambda \mu}(t) $ the operator taking
  ${\cal L}( {\cal H}^1 (\Lambda_n), {\cal H}^{-1}(\Lambda_n))$
into itself and defined by:
  %----
$$\widetilde K _{\lambda \mu}(t) \Big ( \Phi \Big) =  \alpha
_{\Lambda _n}^{(t)}\Big ( K_{\lambda \mu} \big ( \alpha _{\Lambda
_n}^{(-t)}
 \Phi \big ) \Big ) \hskip 1cm \forall \Phi \in
 {\cal L}( {\cal H}^1 (\Lambda_n), {\cal H}^{-1}(\Lambda_n)) \leqno (5.3)$$
%------
With these notations the system becomes
%-------
$$  {d \over dt}  \Phi _{\lambda }^0 (t)  = \Phi _{\lambda }^1 (t)
\hskip 1cm  {d \over dt}  \Phi _{\lambda }^1 (t)  = -i   \sum _{\mu
\in \Lambda _n} \widetilde K _{\lambda \mu}(t) \big ( \Phi_{\mu}^0
(t) \big). \leqno (5.4)$$
%------
Summing up, for all $A$ in ${\cal L} ({\cal H})$,
supported in $\Lambda_n$, the functions $ \Phi_{\lambda }^j
(t)$ defined in (5.2) ($\lambda \in \Lambda_n$) are
$C^1$ from $\R$ to ${\cal L} ( {\cal H}^{1}
\Lambda _n) , {\cal H}^{-1}(\Lambda_n) )$. These maps  are bounded
independently of
$t$ and verify (5.4). It is the unique solution to (5.4) having these
properties together with:
%-----
$$  \Phi_{\lambda }^0 (0) = [ A , Q_{\lambda}] \hskip 1cm
\Phi_{\lambda } ^1(0) = [ A , P_{\lambda}] \leqno (5.5)$$
%------
{\it Second step.} One may find operators-valued matrices
$( A_{\lambda \mu} ^0 (t), A_{\lambda \mu} ^1 (t))$ in ${\cal L}
({\cal L} ({\cal H}))$ satisfying:
%----
$$ {d \over dt} A_{\lambda \mu} ^0 (t) = A_{\lambda \mu} ^1 (t)
\hskip 1cm {d \over dt} A_{\lambda \mu} ^1 (t) = -i  \sum _{\nu \in
\Lambda _n} \widetilde K _{\lambda \nu } (t) \circ A_{\nu \mu} ^0
(t)\leqno (5.6)$$
%----
$$ A_{\lambda \mu} ^0 (0)  = \delta _{\lambda \mu} I  \hskip 1cm
A_{\lambda \mu} ^1 (0)  = 0 \leqno (5.7)$$
%----
In (5.6) the composition is now the composition in ${\cal L}
({\cal L} ({\cal H}))$ and in (5.7) the identity operator is the identity
in
${\cal L} ({\cal L} ({\cal H}))$. Indeed, for all $\gamma $ in
 $]0, \gamma _0[ $, the hypotheses in Proposition B.1 are satisfied, by
Proposition 3.4. If $\gamma$ is in
$]0, \gamma_0[$ and if $M> \sqrt {S_{\gamma}}$ there exists $C(M ,
\gamma)$ such that
%----
$$ \Vert  A_{\lambda \mu} ^j (t) \Vert _{{\cal L} ({\cal L} ({\cal
H}))} \leq C(M, \gamma) e^{- \gamma |\lambda - \mu |  } \leqno
(5.8)$$
%----
We can find, by a similar construction,  operators-valued matrices $( B_{\lambda
\mu} ^0 (t), B_{\lambda \mu} ^1 (t))$ of ${\cal L} ({\cal L}
({\cal H}))$ satisfying the same differential system (5.6) together with
the same estimates (5.8) and the new initial conditions :
%----
$$ B_{\lambda \mu} ^0 (0)  = 0  \hskip 1cm B_{\lambda \mu} ^1 (0)
 = \delta _{\lambda \mu} I. \leqno (5.9)$$
%----
 Suppose now that $A$ belongs to ${\cal W}_1$
and is supported in $\Lambda_n$. The operators $ [A , Q_{\lambda}]$ and $
[A ,
P_{\lambda}]$ are in ${\cal L} ({\cal H})$.  We then define the operators
in ${\cal L} ({\cal H})$ by:
%-----
$$ \Psi _{\lambda }^j (t) = \sum _{\mu \in \Lambda _n}
 A_{\lambda \mu} ^j (t)  \big (  [ A , Q_{\mu} ] \big ) \  + \
 B_{\lambda \mu} ^j (t)  \big (  [ A , P_{\mu} ] \big )
 \hskip 1cm j= 0 , 1$$
 %------
These functions, taking values into ${\cal L} ({\cal H})$, satisfy the same
differential system (5.4) with the same initial conditions (5.5) as the
functions $\Phi _{\lambda}^j
(t)$ (being a priori in
${\cal L} ({\cal H}^1(\Lambda_n) ,{\cal H}^{-1} (\Lambda_n) )$.
Uniqueness shows that  $\Phi _{\lambda}^j (t) = \Psi
_{\lambda}^j (t)$. The functions $\Phi
_{\lambda}^j (t) $ defined in (5.2) have therefore the stated properties.

\hfill \carre

\bigskip
For all $\lambda$ and $\mu$ in $\Lambda_n$ the commutator $[
Q_{\lambda}^{(j)}, \alpha^{(t)} _{\Lambda _n} (Q_{\mu}^{(k)})]$,
 ($0 \leq j, k \leq 1$) is bounded from
 ${\cal H}^1 (\Lambda_n)$ into $ {\cal H}^{-1} (\Lambda_n)$.
 We shall obtain that it is an element of ${\cal L} ({\cal H})$ and we
shall estimate its norm.

\bigskip
\noindent {\bf Proposition 5.2.}  {\it Under the hypotheses (H1)
and (H2)  of section 1, for all $\lambda$ and $\mu$ in
$\Lambda_n$, the  commutator $[ Q_{\lambda}^{(j)}, \alpha^{(t)}
_{\Lambda _n} (Q_{\mu}^{(k)})]$
 ($0 \leq j, k \leq 1$),
is a bounded operator in ${\cal H}$. Moreover, for all
$\gamma$ in $]0, \gamma _0[$ and for all $M>\sqrt
{S_{\gamma}}$, there exists $C(M, \gamma)>0$, (independent of $n$,
$t$, $\lambda$ and $\mu$) such that:
%-----
$$ \Big \Vert \big [Q_{\lambda}^{(j)} , \alpha^{(t)} _{\Lambda _n}
(Q_{\mu}^{(k)}) \big ]\Big \Vert \leq C(M, \gamma)  e^{M|t|} e^{-\gamma d
(\lambda , \mu)}\hskip 1cm 0 \leq j, k \leq 1 $$
%-------
}

\bigskip
\noindent {\it Proof.}  Using the matrices
  $A_{\lambda \mu} ^j (t)$ and   $B_{\lambda \mu} ^j (t)$
  $(j= 0, 1)$ defined in the second step of  Proposition 5.1 one shows
that:
  %----
  $$   [P_{\lambda} , \alpha^{(t)}_{\Lambda _n} ( Q_{\mu}^{(j)}) ]=
  A_{\lambda \mu} ^{(j)} (t) (I)\hskip 1cm
     [Q_{\lambda} , \alpha^{(t)}_{\Lambda _n} ( Q_{\mu}^{(j)}) ]=
  B_{\lambda \mu} ^j (t) (I) \hskip 1cm 0 \leq j \leq 1$$
  %-----
The proof uses the same points as those in  Proposition 5.1.
Then Proposition 5.2  follows from the estimates on these matrices being
analyzed in Proposition B.1.

\hfill \carre

\bigskip

Let us now consider commutators of length two.

\bigskip
\noindent {\bf Proposition 5.3.} {\it If $A$ is in ${\cal
W}_2$, then the commutators $ \big [  [  \alpha ^{(t)} _{\Lambda_n} (A ),
Q_{\lambda_1}^{(j_1)} ], \
Q_{\lambda_2}^{(j_2)} \big ]$ are in ${\cal L}({\cal H})$, ($t\in
\R$, $\lambda_1$ and $\lambda _2$ in $\Lambda_n$, $0 \leq j_1 , j_2
\leq 1$).  Moreover, if $\gamma$ is in $]0, \gamma_0[$ and if $M> 2 \sqrt
{S_{\gamma}}$ there exists $C=C(M, \gamma)$ such that:
%----
$$ \Big \Vert \ \big [  [  \alpha ^{(t)} _{\Lambda_n} (A),
Q_{\lambda_1}^{(j_1)} ], \
Q_{\lambda_2}^{(j_2)} \big ] \ \Big \Vert _{{\cal L}({\cal H})}
\leq C  e^{ M |t|} \Big [ \sum _{(\mu_1 , \mu_2)\in \Lambda_n^2
\atop 0 \leq k_1 , k_2 \leq 1} e^{-\gamma [ |\lambda_1 - \mu_1|+
|\lambda_2 - \mu_2|]} \Big \Vert \ \big [[
 A , Q_{\mu_1}^{(k_1)} ] ,
Q_{\mu_2}^{(k_2)}\big ] \ \Big \Vert \ \ + ... $$
%-----
$$ ... _ +\ \sum _{ \nu \in \Lambda_n \atop 0 \leq k \leq 1} e^{-
\gamma d( \nu , \{  \lambda _1 , \lambda_2\} ) } \big \Vert \ [A ,
Q_{\nu}^{(k)} ] \ \big \Vert  \ \Big  ] $$
%----

}

\bigskip
\noindent {\it First step.} Set $A$ in ${\cal L}({\cal
H})$. We show that the functions defined for all real $t$ by:
%----
$$ \Phi _{\lambda_1 \lambda_2} ^{j_1,j_2} (t) =  \big [  [A ,
\alpha ^{(t)} _{\Lambda_n} (Q_{\lambda_1}^{(j_1)}) ], \ \alpha
^{(t)} _{\Lambda_n} (Q_{\lambda_2}^{(j_2)})\big ] \hskip 1cm 0
\leq j_1 , j_2 \leq 1  \leqno (5.10)$$
%------
are $C^1$ functions taking values from $\R$ into ${\cal
L}({\cal H}^2 (\Lambda_n), {\cal H}^{-2} (\Lambda_n))$ and
verifying the following differential system where the
operators $\widetilde K _{\lambda \mu} (t) $ are defined in
(5.3) and where the operators $R_{\lambda \mu }$ are given by Proposition
3.5:
%----------
%----
$$ {d\over dt}  \Phi _{\lambda_1 \lambda_2} ^{00} (t) =  \Phi
_{\lambda_1 \lambda_2} ^{01} (t) +  \Phi _{\lambda_1 \lambda_2}
^{10} (t)\leqno (5.11) $$
%------
 %----
  $$ {d\over dt}  \Phi _{\lambda_1 \lambda_2} ^{10} (t)\ =\
  \Phi _{\lambda_1 \lambda_2}^{11} (t)
  -i  \sum _{\mu_1 \in \Lambda_n} \widetilde K _{\lambda_1 \mu_1} (t)
\big (  \Phi _{\mu_1 \lambda_2} ^{00} (t)\big ) \leqno (5.12)$$
%----
%------
$$ {d\over dt}  \Phi _{\lambda_1 \lambda_2} ^{01} (t)\ =\
  \Phi _{\lambda_1 \lambda_2}^{11} (t)
  -i  \sum _{\mu_2 \in \Lambda_n}  \widetilde K _{\lambda_2 \mu_2} (t)
\big (  \Phi _{\lambda_1 \mu_2} ^{00} (t)\big ) \leqno (5.13) $$
%--------
 %---
  $$ {d\over dt} \Phi _{\lambda_1 \lambda_2} ^{11} (t)\ = -i  \sum _{\mu_1 \in
\Lambda_n} \widetilde K _{\lambda_1 \mu_1} (t) \big ( \Phi _{\mu_1
\lambda_2} ^{01} (t) \big )\ -i  \ \sum _{\mu_2 \in \Lambda_n}
 \widetilde K _{\lambda_2 \mu_2} (t)
\big ( \Phi _{\lambda_1 \mu_2} ^{10} (t) \big )\ +\ F_{\lambda _1
, \lambda_2} (t) \leqno (5.14)$$
 %-----
 $$ F_{\lambda _1 , \lambda_2} (t) = -  \sum _{\mu_1 \in \Lambda_n}
 \alpha ^{(t)}
_{\Lambda_n} \Bigg ( R_{\lambda_1 \mu_1}
 \Big ( [\alpha ^{(-t)} _{\Lambda_n} (A) , Q_{\mu_1} ]\ , \
  P_{\lambda_2} \Big ) \Bigg )  \leqno (5.15)$$
%----
The system of   functions $\Phi _{\lambda_1 \mu_2} ^{10} (t)$ is the
unique solution to the differential system  (5.11)... (5.15) satisfying
the initial conditions:
%-----
$$ \Phi _{\lambda_1 \lambda_2} ^{j_1j_2} (0)\ = \  \big [  [A ,
Q_{\lambda_1}^{(j_1)} ], \ Q_{\lambda_2}^{(j_2)}\big ]
 \hskip 1cm 0 \leq j_1 , j_2\leq 1 \leqno (5.16) $$
%---

Let us give more details, says, for the proof of (5.14).
Following the differential system satisfied by $ \alpha
^{(t)} _{\Lambda_n} (Q_{\lambda})$  and $ \alpha ^{(t)}
_{\Lambda_n} (Q_{\mu})$, (see the first step of Proposition
4.2) one observes that:
%----
 $$ {d\over
dt} \Phi _{\lambda_1 \lambda_2} ^{11} (t)\ =\ - i \ \alpha ^{(t)}
_{\Lambda_n} \Bigg ( \Big [ \ \big [ \alpha ^{(-t)} _{\Lambda_n} (A) ,
[P_{\lambda_1}, V_{\Lambda_n}]\ \big  ] , P_{\lambda_2} \Big ] + \Big [
[\alpha ^{(-t)} _{\Lambda_n} (A) , P_{\lambda_1}
 ] \ , \ [P_{\lambda_2} , V_{\Lambda_n}] \Big ]
 \Bigg)$$
%-------
Using the operators $K_{\lambda \mu}$
 of  proposition 3.3, one gets:
%------
$$  \Big [ [\alpha ^{(-t)} _{\Lambda_n} (A) , P_{\lambda_1}
 ] , [P_{\lambda_2} , V_{\Lambda_n}] \Big ] =
  \sum _{\mu_2 \in \Lambda_n}\ K_{\lambda_2 \mu_2}
  \Big ( \big [  [\alpha ^{(-t)} _{\Lambda_n} (A) ,
   P_{\lambda_1} ]  , Q_{\mu_2} \big ] \Big )  $$
%-------
Also using the operators $R_{\lambda \mu}$
 of proposition 3.5, one sees that:
$$ \Big [\ \big  [\alpha ^{(-t)} _{\Lambda_n} (A) , [P_{\lambda_1},
V_{\Lambda_n}]\ \big  ] , P_{\lambda_2} \Big ]  =
 \sum _{\mu_1 \in \Lambda_n}\ K_{\lambda_1 \mu_1}
 \Big ( \Big [ \ [\alpha ^{(-t)} _{\Lambda_n} (A) , Q_{\mu_1} ] ,
 , P_{\lambda_2} \Big ] \Big ) + R_{\lambda_1 \mu_1}
 \Big ( [\alpha ^{(-t)} _{\Lambda_n} (A) , Q_{\mu_1} ]\ , \
  P_{\lambda_2} \Big ) $$
  %-------
Equalities (5.14) and (5.15) then follows.

 \bigskip \noindent {\it Second step.}  Suppose now that $A$ is in ${\cal
W}_2$. We shall show that the operators $F_{\lambda _1 , \lambda_2} (t)$
 defined in (5.15) are in ${\cal L}({\cal H})$ and we shall estimate their
norms. More precisely, we shall show that if
 $\gamma \in ]0 , \gamma_0[ $ and $M> \sqrt {S_{\gamma}}$, one has:
 %----
 $$ \Vert F_{\lambda _1 , \lambda_2} (t) \Vert_{{\cal L}({\cal H})}
\leq C e^{M|t|} \sum _{ \nu \in \Lambda_n \atop 0 \leq k \leq 1}
e^{-\gamma_0 |\lambda_1 - \lambda_2| - \gamma {\rm dist} (\nu , \{
\lambda_1 , \lambda_2 \} ) }
 \Big \Vert \  [A , Q_{\nu}^{(k)} ] \ \Big \Vert \leqno (5.17)$$
%-----
Indeed, from Proposition 3.5, if $\lambda_1 \not=
\lambda_2$,
 then the sum in (5.15) is reduced to two terms: the one with $\mu_1 =
\lambda_1$ together with the one with $\mu_1 = \lambda_2$. In this case,
one has:
%----
$$ \Vert  F_{\lambda _1 , \lambda_2} (t)\Vert_{{\cal L}({\cal H})}
\leq C e^{-\gamma_0 |\lambda _1 - \lambda_2|} \Big [  \Vert[\alpha
^{(-t)} _{\Lambda_n} (A) , Q_{\lambda_1} ] \Vert_{{\cal L}({\cal
H})} +  \Vert[\alpha ^{(-t)} _{\Lambda_n} (A) , Q_{\lambda_2} ]
\Vert_{{\cal L}({\cal H})} \Big ] $$
%-----
If $\lambda_1 = \lambda_2$, one has, from Proposition 3.5:
%----
$$ \Vert  F_{\lambda _1 , \lambda_1} (t)\Vert_{{\cal L}({\cal H})}
\leq C  \sum _{\mu_1 \in \Lambda_n}\  e^{-\gamma_0
|\lambda_1-\mu_1|}  \Vert[\alpha ^{(-t)} _{\Lambda_n} (A) ,
Q_{\mu_1} ] \Vert_{{\cal L}({\cal H})}$$
%---
Following Proposition 5.1, one sees, if $M> \sqrt{S_{\gamma}}$:
%-----
$$ \Vert[\alpha ^{(-t)} _{\Lambda_n} (A) , Q_{\mu_1} ]
\Vert_{{\cal L}({\cal H})}  =  \Vert[ A ,\alpha ^{(t)}
_{\Lambda_n} (Q_{\mu_1}) ] \Vert_{{\cal L}({\cal H})} \ \leq C(M,
\gamma) e^{ M|t|} \sum _{\nu \in \Lambda_n \atop 0 \leq  k \leq 1}
e^{-\gamma| \mu_1 - \nu|} \Vert [A , Q_{\nu}^{(k)} \Vert_{{\cal
L}({\cal H})}$$
%-------
and the estimates (5.17) are easily deduced.

 \bigskip \noindent {\it Third step.}
 If $A$ is in ${\cal W}_2$,  then the initial data  (5.16) are
 in ${\cal L}({\cal H})$. From the remarks below Proposition B.1, if
$\gamma$ is in
 $]0, \gamma_0[$, the system
 (5.11)...(5.14) has a solution $ \Psi _{\lambda_1 \lambda_2} ^{j_1j_2}
 (t)$ in ${\cal L}({\cal H})$ satisfying (5.16). Moreover, if $M>
 2 \sqrt {S_{\gamma}}$, there exists $C(M, \gamma)$ such that:
 %-----
 $$ \Vert  \Psi _{\lambda_1 \lambda_2} ^{j_1j_2}(t)\Vert _{{\cal L}({\cal
 H})} \ \leq C(M, \gamma)e^{M|t|}  \sum _{(\mu_1 , \mu_2) \in \Lambda_n^2
 \atop 0 \leq k_1 , k_2 \leq 1 }e^{-\gamma (|\lambda_1 - \mu_1| +
  |\lambda_2 - \mu_2|)} \Vert \big [  [A ,
Q_{\mu_1}^{(k_1)} ], \ Q_{\mu_2}^{(k_2)}\big ] \Vert _{{\cal
L}({\cal  H})} + ...$$
 %----
$$ ... + C(M, \gamma)  \sum _{(\mu_1 , \mu_2) \in \Lambda_n^2}
e^{-\gamma (|\lambda_1 - \mu_1| +
  |\lambda_2 - \mu_2|)} \int_0^t e^{M|t-s|}
  \Vert F_{\mu _1 , \mu_2} (s) \Vert _{{\cal
L}({\cal  H})} ds$$
%----
The proof of this proposition then follows from the estimates of  $F_{\mu
_1 , \mu_2} (s) $ in (5.17).
%----

\bigskip
\noindent {\bf 6. Evolution for a finite number of sites.}
\bigskip
From proposition 4.1, the operator $e^{itH_{\Lambda
_n}}\otimes I $ is bounded in the ${\cal H}^k(\Lambda_n)$.
However, when following the proof of Proposition 4.1 the norm of this
operator could depend on
 $n$. On the contrary, the next proposition provides a bound independent
on $n$.

\bigskip
\noindent {\bf Proposition  6.1.} {\it  The operator
$e^{itH_{\Lambda _n}} \otimes I$ is bounded  in ${\cal H}^k$,
($0 \leq k\leq 2$) with a norm $\leq C_k e^{M_k|t|} $ where $C_k>0$
and $M_k>0$ are independent of all the parameters.  For all $A\in {\cal L}
({\cal H}^k, {\cal H}) $, ($k=1, 2$) with finite support,  if
$\Lambda _n$ contains the support of $A$, one has:
%----
$$ \Vert \alpha ^{(t)} _{\Lambda _n}(A)\Vert _{{\cal L}({\cal H}^k
, {\cal H})} \leq C_k e^ {M_k |t|}  \Vert  A \Vert _{{\cal
L}({\cal H}^k , {\cal H})} \leqno (6.1)$$
%----
}

\bigskip
\noindent {\it Proof.}  Set $f\in {\cal H}^1$.
From Proposition 4.2 and for all $\lambda \in \Lambda_n$ one see:
%----
$$ \Vert Q_{\lambda} (e^{itH_{\Lambda _n}} \otimes I)f \Vert =
\Vert \alpha^{(-t)}_{\Lambda_n} (Q_{\lambda} ) f \Vert \leq \Vert
R_{\lambda} (-t) f \Vert + \sum _{\mu \in \Lambda_n} \Big [\Vert
A_{\lambda \mu}^{(n)} (-t) Q_{\mu} f \Vert + \Vert B_{\lambda
\mu}^{(n)} (-t) P_{\mu} f \Vert \Big ]$$
%-------
We deduce from the estimates (4.5) and (4.6) that, if
$\gamma \in ]0, \gamma_0[$ and if $M_1> \sqrt {S_{\gamma}}$ then
%----
$$  \Vert Q_{\lambda} (e^{itH_{\Lambda _n}} \otimes I)f \Vert\
\leq \  C_1 e^{M_1|t|} \Vert f \Vert _{{\cal H}^1} $$
%-----
with $C_1>1$ independent of $n$ and $t$. If $\lambda $ is not in
$\Lambda_n$ then the same inequality is valid since
$Q_{\lambda}$ commutes with $e^{itH_{\Lambda _n}} \otimes I$. We proceed
similarly  with the operators $P_{\lambda}$ proving that
 $\Vert e^{itH_{\Lambda _n}} \otimes I \Vert _{ {\cal L}
( {\cal H}^1 )} \leq C_1 e^{M_1|t|} $.

\smallskip
\noindent {\it Action in ${\cal H}^2$.} For all $\lambda_1$ and
$\lambda_2$ in $\Lambda _n$ we have from the above points:
%---
$$ \Vert  Q_{\lambda_1}^{(j_1)} \
 Q_{\lambda_2}^{(j_2)}  ( e^{itH_{\Lambda_n}}\otimes I)  f \Vert =
 \Vert  Q_{\lambda_1}^{(j_1)} \  ( e^{itH_{\Lambda_n}}\otimes I)
 \alpha ^{(-t)} _{\Lambda _n}( Q_{\lambda_2}^{(j_2)} )  f \Vert
  \leq C_1 e^{M_1|t|} \Vert
 \alpha ^{(-t)} _{\Lambda _n}( Q_{\lambda_2}^{(j_2)} ) f \Vert_{{\cal
H}^1} \ $$
%-----
One see:
%-----
$$ \Vert Q^{(k) }_{\mu}  \alpha ^{(-t)} _{\Lambda _n}(
Q_{\lambda_2}^{(j_2)} ) f \Vert \ \leq \ \Big \Vert  \big [ Q^{(k)
}_{\mu}\ ,\   \alpha ^{(-t)} _{\Lambda _n}( Q_{\lambda_2}^{(j_2)}
)\big ]   f\Big  \Vert \ + \ \Vert  \alpha ^{(-t)} _{\Lambda _n}(
Q_{\lambda_2}^{(j_2)} )\  Q^{(k) }_{\mu}  f \Vert$$
%----
$$ \leq  C'_1 e^{M_1 |t|} \Big [\  \Vert f \Vert +\ \Vert Q^{(k)
}_{\mu} f \Vert_{{\cal H}^1}\ \Big ]$$
%----
for all $\mu \in \Lambda_n$.

The two above terms have been estimated using
Propositions 5.2 and 5.1 respectively.  One deduces (with another constant
$C_2$) that, $\Vert Q_{\lambda_1}^{(j_1)} \
 Q_{\lambda_2}^{(j_2)} ( e^{-itH_{\Lambda_n}} \otimes I) f \Vert
 \leq C_2 e^{2 M_1 |t|}\Vert f \Vert_{{\cal H}^2}$.  The proof is
completed.

  \hfill \carre

\bigskip

\bigskip
\noindent {\bf Theorem 6.2.} {\it  If $A$ is in ${\cal
W}_k$ with a finite support,  and if $\Lambda _n$ contains the support of
$A$, then $\alpha ^{(t)} _{\Lambda _n}(A)$ is in ${\cal W}_k$ ($0
\leq k \leq 2$). Moreover, there exists two constants $C_k$ and
$M_k$ independent of  $A$,  $n$ and of $t$, such that:
%----
$$\Vert \alpha ^{(t)} _{\Lambda _n} (A) \Vert _{{\cal W}_k} \leq
C_k  e^{M_k|t|}   \Vert A \Vert _{{\cal W}_k} \leqno (6.2)$$
%----

}

\bigskip
\noindent {\it Proof.} The norm in ${\cal L} ({\cal
H})$ is conserved by $ \alpha ^{(t)} _{\Lambda _n}$. By Proposition 5.1,
if $A\in {\cal W}_1$ is supported in $\Lambda
_n$ and if $\lambda \in \Lambda_n$ then the commutators of $A$ with
 $\alpha_{\Lambda_n}^{(-t)}(Q_{\lambda}^{(j)})$ are bounded operators.
Thus, the commutators of
 $\alpha_{\Lambda_n}^{(t)}(A)$ with $Q_{\lambda}^{(j)} $ are bounded
operators if $\lambda \in
 \Lambda _n$. Since these commutators are vanishing when $\lambda \notin
 \Lambda_n$ then $\alpha_{\Lambda_n}^{(t)}(A)$ is in ${\cal W}_1$.
 If $\gamma >0$ is in $]0, \gamma_0[$, and if $M_1> \sqrt {
 S_{\gamma}}$, we see that:
 %---
 $$ \sum _{\lambda \in \Lambda _n \atop 0 \leq j \leq 1} \Vert \
 [\alpha_{\Lambda_n}^{(t)}(A), Q_{\lambda}^{(j)} ] \ \Vert \leq
 C(M_1, \gamma ) \  e^{M |t|}\  \sum _{(\lambda ,\mu )\in
 \Lambda_n^2 \atop 0 \leq j, k \leq 1} e^{-\gamma d( \lambda , \mu)}
\Vert \
 [A, Q_{\mu}^{(k)} ] \ \Vert $$
 %----
$$ \leq  C_1(M_1, \gamma ) \  e^{M |t|} \Vert A \Vert _{{\cal
W}_1}
 \sup _{\mu \in \Z}
\sum _{\lambda \in \Z}  e^{-\gamma d( \lambda , \mu)} $$
%-----
Consequently, there are $C_1>0$ and $M_1>0$ such that (6.2) is valid for
$k=1$.
%---
\smallskip
\noindent {\it Action in ${\cal W}^2$.}   Proposition 5.3
shows that the commutators written as  $[ \alpha ^{(t)} _{\Lambda
_n}(A) , Q_{\lambda_1}^{(j_1)} ], Q_{\lambda_2}^{(j_2)}]$ are bounded
operators and are vanishing if $\lambda_1$ or $\lambda_2$ is not in
$\Lambda_n$. Consequently, $ \alpha ^{(t)} _{\Lambda
_n}(A)$ is in ${\cal W}_2$. If $\gamma$
 is in $]0, \gamma_0[$ and $M_2> 2 \sqrt {S_{\gamma}}$ then
 Proposition 5.3 implies that inequality (6.2) is verified for $k=2$.

  \hfill \carre

\bigskip

\bigskip

\noindent {\bf 7. Existence of  dynamics in the
Weyl algebra. }
\medskip
The number of sites shall now goes to infinity. The proofs of theorem 1.1
and 1.2 on the existence
of a limit  rely on the description  of the difference
 $\alpha ^{(t)} _{\Lambda _m}(A) -
\alpha ^{(t)} _{\Lambda _n}(A)$.

\bigskip
\noindent {\bf Proposition 7.1.} {\it There exists $C>0$, $M>0$ and
$\gamma
>0$ satisfying the following properties. For all $A\in {\cal W}_2$
with finite support and for all integers $m$ and $n$
verifying  $0 < m < n$ and such that $\Lambda _m$ contains the support
$\sigma (A)$ of $A$,  for all $t\in \R$, one has:
%----
$$ \Vert \alpha ^{(t)} _{\Lambda _m}(A) - \alpha ^{(t)} _{\Lambda
_n}(A) \Vert _{{\cal L} ( {\cal H}^1 , {\cal H}^0)} \leq C  \Vert
A \Vert _{{\cal W}_2}\  e^{M|t|}  e^{- \gamma d( \sigma(A),
\Lambda_m^c)}  \leqno (7.1)$$
%------
}

\bigskip

%-----HOMOTOPIE--------

\noindent {\it Proof.} For $m<n$ we denote by $V^{inter
}_{mn}$ the potential of the interaction between $\Lambda _m$
and $\Lambda _n \setminus \Lambda_m$:
%----
$$ V^{inter}_{mn} (x) = -b Q_m Q_{m+1} \ -  b Q_{-m} Q_{-m-1} +
\sum _{ (\lambda , \mu) \in E_{mn}} V_{\lambda \mu}$$
%-------
where $E_{mn}$ denotes the set of pairs of sites $(\lambda , \mu)$
such that, one of the site ($\lambda$ or $\mu$) is in $ \Lambda
_m$ and the other site belongs to $ \Lambda _n \setminus \Lambda _m$.
For all $\theta \in [0, 1]$, set:
%-----
$$ H_{mn \theta } = H _{\Lambda_n} - (1- \theta) V^{inter}_{mn} $$
%----
One may define a unitary operator by $e^{i t H_{mn\theta}}$ and
set:
%---
$$\alpha ^{(t)} _{m n \theta } (A)= (e^{i t H_{mn\theta}}\otimes
I) \  A \ (e^{-i t H_{mn\theta}}\otimes I)$$
%--------
Thus, if $A$ is supported  in $\Lambda _m$ and if $m<n$:
%----
 $$\alpha ^{(t)} _{m n 1 } (A) = \alpha^{(t)} _{\Lambda_n}
(A)  \hskip 1cm \alpha ^{(t)} _{m n 0 } (A) = \alpha^{(t)}
_{\Lambda_m} (A) $$
%---
The function $\varphi (t , \theta)= {\partial \over \partial
\theta} \alpha^{(t)}_{mn\theta} (A)$ verifies:
%------
$$ {\partial \varphi \over \partial t} = i \big [ H_{mn \theta } ,
\varphi \big ]\ + \ i \ \big [ V_{mn}^{inter} \ ,\
\alpha^{(t)}_{mn\theta} (A)\ \big ] \hskip 1cm \varphi ( 0,
\theta) = 0$$
%---
Consequently:
%----
$$ { \partial \over \partial \theta  }  \alpha ^{(t)} _{m n \theta
} (A)= i \int_0^t \alpha ^{(t- s)} _{m n \theta } \big ( [
V^{inter}_{mn} ,  \alpha ^{(s)} _{m n \theta } (A) ] \big ) ds$$
%----
One obtains the integral representation:
%-----
$$ \alpha^{(t)} _{\Lambda_n} (A) - \alpha^{(t)} _{\Lambda_m} (A) =
i \int _0 ^t \int _0^1 \alpha ^{(t- s)} _{m n \theta } \big ( [
V^{inter}_{mn} ,  \alpha ^{(s)} _{m n \theta } (A) ] \big ) ds
d\theta$$
%------
Applying Proposition 6.1 to the operator $H_{mn
\theta}$ which verifies the same hypotheses as
$H_{\Lambda_n}$, we deduce that there exists $C>0$ and $M>0$ such that:
%------
$$ \Vert   \alpha^{(t)} _{\Lambda_n} (A) - \alpha^{(t)}
_{\Lambda_m} (A)     \Vert_{{\cal L} ( {\cal H}^1 , {\cal H}^0)}
\leq C  \int _0 ^t \int _0^1  e^{M|t-s|}  \Vert  [ V^{inter}_{mn}
,  \alpha ^{(s)} _{m n \theta } (A) ] \Vert_{{\cal L} ( {\cal H}^1
, {\cal H})} ds d\theta $$
%----
 for all $(\lambda , \mu)$ in
$E_{mn}$. Applying Proposition 3.2 to the operator
 $\alpha ^{(s)} _{m n \theta } (A)$ belonging in ${\cal W}_2$ we obtain:
 %-----
 $$ \Vert  [ V_{\lambda \mu} ,  \alpha ^{(s)} _{m n \theta } (A) ]
\Vert_{{\cal L} ( {\cal H}^1 , {\cal H}^0)} \ \leq C e^{-\gamma_0
|\lambda - \mu|}\  \sum _{ 1 \leq j+k \leq 2} \ \Vert   (ad\
Q_{\lambda})^j  \ (ad\ Q_{\mu})^k \  \alpha ^{(s)} _{m n \theta }
(A)
 \Vert \ $$
 %-----
Similarly:
%------
 $$ \Vert  [ Q_m Q_{m+1} ,  \alpha ^{(s)} _{m n \theta } (A) ]
\Vert_{{\cal L} ( {\cal H}^1 , {\cal H}^0)} \ \leq C \sum _{ 1
\leq j+k \leq 2} \ \Vert   (ad\ Q_{m})^j  \ (ad\ Q_{m+1})^k \
\alpha ^{(s)} _{m n \theta } (A)
 \Vert \ $$
 %-----
Summing on the pairs $(\lambda , \mu)$ in $E_{mn}$ we get:
%-----
 $$ \Vert  [ V^{inter}_{mn} ,  \alpha ^{(s)} _{m
n \theta } (A) ] \Vert_{{\cal L} ( {\cal H}^1 , {\cal H}^0)} \leq
C \sum _{(\lambda , \mu) \in E_{mn}} e^{-\gamma_0 |\lambda - \mu|}
\sum _{ 1 \leq j+k \leq 2}
 \Vert   (ad\ Q_{\lambda})^j  \ (ad\ Q_{\mu})^k \  \alpha ^{(s)} _{m n
\theta } (A)
 \Vert \ $$
 %-----
Consequently:
 %--------
 $$ \Vert   \alpha^{(t)} _{\Lambda_n} (A) - \alpha^{(t)}
_{\Lambda_m} (A)     \Vert_{{\cal L} ( {\cal H}^1 , {\cal H}^0)}
\leq ... \leqno (7.2)$$
%---
$$ ... \leq \ C  \sum _{(\lambda , \mu) \in E_{mn} \atop  1 \leq
j+k \leq 2 } e^{-\gamma_0 |\lambda - \mu|} \int _0 ^t \int _0^1
e^{M|t-s|} \Vert   (ad\ Q_{\lambda})^j  \ (ad\ Q_{\mu})^k \ \alpha
^{(s)} _{m n \theta } (A)
 \Vert \ ds d\theta  $$
%---
\bigskip

Proposition 7.1 then follows from the next lemma,  which shall also be
used in section 8.

 \bigskip
\noindent {\bf Lemma 7.2.}  {\it If $\gamma$ is in $]0, \gamma_0[$
and if $M> 2 \sqrt { S_{\gamma}}$ then there exists $C(M, \gamma)$ such
that,
for all $n$, for all disjoint sets $E_1$ and $E_2$ included
in $\Lambda _n$, for all $A\in {\cal W}_2$ supported in
$E_1$, we have:
%----
$$ \sum _{(\lambda_1 , \lambda _2) \in \Lambda _n \times E_2 \atop
1 \leq \alpha + \beta  \leq 2 }
e^{-\gamma_0 |\lambda_1 - \lambda_2|} \Vert
(ad\  Q_{\lambda_1})^{\alpha}  \
(ad\ Q_{\lambda_2})^{\beta} \  \alpha ^{(s)} _{m n \theta } (A)
 \Vert  \leq C(M, \gamma)  \  \Vert A \Vert
_{{\cal W}_2}\  e^{M|s|}  e^{- \gamma d(E_1 , E_2) }$$
%------
}

\bigskip

This lemma is deduced from  propositions 5.1 and 5.3 applied to
the Hamiltonian $H_{mn\theta}$. Proposition 7.1 is a consequence of
(7.2) together with this lemma,  setting $E_1 = \sigma (A)$ and $E_2 =
\Lambda _n \setminus \Lambda _m$.

\bigskip

\noindent {\it Proof of theorem 1.1 and 1.2.} From
Proposition 7.1 the sequence $\alpha ^{(t)} _{\Lambda _n}(A)$ is a
Cauchy sequence in ${\cal L} ({\cal H}^2 , {\cal H})$ and thus converges
in ${\cal L} ({\cal H}^2 , {\cal H})$ towards an element which is noted
$\alpha ^{(t)} (A)$. By Proposition 6.2, we have
$\Vert \alpha ^{(t)} _{\Lambda _n}(A) \Vert _{{\cal W}_2} \leq C
e^{M|t|} \Vert A \Vert _{{\cal W}_2}$. Following theorem
1.4 the operator $\alpha ^{(t)} (A)$ is in ${\cal W}_2$ with a
norm $\leq C e^{M|t|} \Vert A \Vert _{{\cal W}_2}$ and for all
$f \in {\cal H}$, the sequence $\alpha ^{(t)} _{\Lambda _n}(A)f$
strongly converges to $\alpha ^{(t)} (A)f$.  The classical continuity
of the  map $t \rightarrow \alpha ^{(t)}_{\Lambda
_n} (A)f$ for all $n$ and for all $f$ together with the above
inequalities, show the continuity of the
map $t \rightarrow \alpha ^{(t)}_{\Lambda _n} (A)f$.

\bigskip

\noindent {\it Extension of $\alpha^{(t)}$ to the algebra ${\cal
W}_2$.} Set  $A$ in ${\cal W}_2$ with an arbitrary support.
From theorem 1.3 there exists a sequence $(A_n)$ in
${\cal W}_2$ with finite supports such that:
%--
$$ \Vert A_n \Vert_{{\cal W}_2} \leq  \Vert A \Vert_{{\cal W}_2}
\hskip 1cm  \lim _{n \rightarrow \infty}
 \Vert A_n -A \Vert_{ {\cal L}( {\cal H}^2 , {\cal
H})}=0$$
%----
The operator $\alpha ^{(t)} (A_n)$ is well-defined in view of
theorem 1.1 and 1.2 since the  $A_n$ have finite support and one has:
%---
$$  \Vert \alpha^{(t)} (A_n) \Vert_{{\cal W}_2} \leq
 C e^{M|t|}  \Vert A_n \Vert_{{\cal W}_2}
  \leq  C e^{M|t|}   \Vert A \Vert_{{\cal W}_2} \leqno (7.3)$$
  %----
  If $m<n$ then we also see from theorem 1.2:
  %----
$$  \Vert \alpha^{(t)} (A_n - A_m) \Vert_{ {\cal L}( {\cal H}^2 ,
{\cal H})} \leq  C e^{M|t|}   \Vert  A_n - A_m \Vert_{ {\cal L}(
{\cal H}^2 , {\cal H})}$$
%----
The sequence $ \alpha^{(t)} (A_n )$ thus converges in $ {\cal L}(
{\cal H}^2 , {\cal H})$ to an element that is denoted  $
\alpha^{(t)} (A )$. From (7.3) and theorem 1.4 this element is  in ${\cal
W}_2$ and it verifies:
%---
$$\Vert  \alpha^{(t)} (A ) \Vert _{{\cal W}_2} \leq
 C e^{M|t|}  \Vert   A  \Vert _{{\cal W}_2}$$
%----
The group $ \alpha^{(t)} $ is therefore extended to the whole algebra
${\cal W}_2$.

\vfill \eject
 \noindent  {\bf 8.  Lieb-Robinson's inequalities.}
\bigskip
\noindent {\bf Proposition 8.1.} {\it For all $\gamma $ in
$]0, \gamma_0[$ and for all $M> 2 \sqrt{ S_{\gamma}}$,
there exists $C (M, \gamma )>0$  such that,  for all $A$
 and $B$ in ${\cal W}_2$ with finite supports $\sigma (A)$ and $\sigma
(B)$,
 for all $n$ such that $\Lambda _n$ contains $\sigma (A)$ and
$\sigma (B)$, for all $t\in \R$ we have:
%------
$$\Big \Vert \ [ \alpha ^{(t)}_{\Lambda _n} (A), B ] \ \Big \Vert
_{{\cal L}( {\cal H}^2 , {\cal H})}  \leq \ C(M, \gamma ) \ \Vert
A \Vert _{{\cal W}_2} \ \Vert B \Vert _{{\cal W}_2}\ e^{M|t|} e^{
- \gamma d ( \sigma (A) , \sigma (B)) } \leqno (8.1)$$
%-----
The same inequality is valid when replacing $ \alpha
^{(t)}_{\Lambda _n}$ by $ \alpha ^{(t)}$.

}

\bigskip
\noindent {\it Proof.} From corollary 2.6 applied
with the  operators $B$ and $\alpha ^{(t)} _{\Lambda _n  } (A)$, (both
having their support in $\Lambda _n$) one has:
 %------
$$ \Vert [ \alpha ^{(t)} _{\Lambda _n  } (A)\ , \ B ] \Vert_{{\cal
L}({\cal H}^2 , {\cal H})} \leq  \ C \  \Vert B \Vert _{{\cal
W}_2}\  \sum _{\lambda \in \sigma (B)\atop 1 \leq j+k \leq 2 }
\Vert (ad \ P_{\lambda})^j (ad \ Q_{\lambda})^k \big (\alpha
^{(t)} _{\Lambda _n } (A) \big )\Vert  $$
%-----
Inequality (8.1) then follows by applying Lemma 7.2
to the sets $E_1 = \sigma (A)$ and $E_2 = \sigma (B)$.
The analogous inequality for  $\alpha ^{(t)} (A)$ is then deduced since
 $\Vert  \alpha ^{(t)} _{\Lambda _n  } (A) -  \alpha ^{(t)}
 (A) \Vert_{ {\cal L}( {\cal H}^2 ,
{\cal H})} $ tends to 0.

\hfill \carre
\bigskip

\noindent {\it Propagation speed.}  Set:
%----
$$ V_0 = \inf _{ 0 < \gamma < \gamma _0} { 2 \sqrt { S_{\gamma}}
\over \gamma} \leqno (8.2)$$
%------
where $S_{\gamma}$ is the constant given in Proposition 3.4. For the case
of interaction with nearest neighbors the infimum bound is taken on  $]0,
\infty[$.

\bigskip
\noindent {\it Proof of theorem 1.5.} Set $A$ and $B$ in
${\cal W}_2$ with finite supports $\sigma (A)$ and $\sigma (B)$.
Set $(h_n , t_n)$ a sequence in $\Z \times \R$ with $|t_n|
\rightarrow \infty$ and with $|h_n| \geq v_1 |t_n|$ where $v_1 > V_0$,
$V_0$ being defined above.  Set $\gamma \in ]0, \gamma_0[$ such that
$2 \sqrt {S_{\gamma}}  < v_1 \gamma$. Set $M$ such that $2 \sqrt
{S_{\gamma}} < M  < v_1 \gamma$. The sequence $M |t_n| - \gamma d(
\sigma (A), ( \sigma ( \tau _{h_n} (B))) $ tends to $-\infty$.
For all  $f\in {\cal H}^2$ the inequality (8.1) (with $\alpha
^{(t)} _{\Lambda _n  }$ replaced with $\alpha ^{(t)}$) shows that:
%-----
$$ \lim _{n \rightarrow \infty} \Big \Vert \ [ \alpha ^{(t_n)}
(A), \tau _{h_n}(B) ]f \ \Big \Vert_{\cal H} = 0$$
%-----
The result is extended by density to all $f\in {\cal H}$.

\hfill  \carre

%------------APPENDICE------AAAAAAA----------------
%----------------APPENDICE-A--------------------

 \noindent {\bf Appendice A.  Approximation of operators. Proof of
Proposition 2.5.}
\bigskip

We shall first prove  Proposition 2.5 for two
finite subsets $E$ and $F$ de $\Z$ such that $E\subset F$ with their
difference $F \setminus E$ being reduced to only one
element $\lambda$.  Operators in ${\cal L}({\cal
H}_F)$ shall be identified using the map $i_{F\Z}$ with the elements of
${\cal L} ({\cal H})$ supported in $F$. We denote by
 ${\cal W}_k (F)$ the set of all $A$ in ${\cal L}({\cal
H}_F)$ such that $i_{F\Z}(A)$ is in ${\cal W}_k$.

\bigskip
\noindent {\bf Proposition A.1.} {\it There exists a constant
$C>0$ such that, for all finite subset $E$ and $F$ in
$\Z$ written as $F = E \bigcup \{ \lambda \}$ where $\lambda \in
\Z \setminus E$, for all
 $T $ in ${\cal W}_2(F)$,
 %-----
 $$ \Vert ( T - i_{EF} \circ \rho_{FE} (T)) f
  \Vert_{ {\cal L}( {\cal H}^2 , {\cal H})} \leq C
\sum _{ 1 \leq j+k \leq 2 }
 \Vert (ad P_{\lambda})^j (ad Q_{\lambda})^k T \Vert _{{\cal L}({\cal
 H})}\ \ \leqno (A.1)$$
%-----

}

\bigskip

\noindent {\it End of the proof of proposition 2.5.} If $E
\subset F \subset G$ then one has $\rho_{ GE} = \rho _{FE} \circ
\rho_{GF}$ and $i_{EG}= i_{FG} \circ i_{EF}$. Consequently, if
$F = E \bigcup \{ \lambda_1 , ... \lambda_m \}$ then we successively apply
proposition A.1 with the set $E_k = E \bigcup
\{ \lambda_1 , ... \lambda_k \}$ ($1 \leq k \leq m$) and $E_0 =
E$. We obtain, for all $T$ in  ${\cal W}_2(F)$:
%-----
$$ \Vert ( T - i_{EF} \circ \rho_{FE} (T))  \Vert_{{\cal L}( {\cal
H}^2 , {\cal H})} \leq \sum _{k=1}^m \Vert T_k
-i_{E_{k-1}E_k}\circ  \rho _{E_k E_{k-1}} ( T_k) \Vert_{{\cal L}(
{\cal H}^2 , {\cal H})} \hskip 1cm T_k = \rho _{F E_k} (T).$$
%-----
Proposition 2.5 thus follows from proposition A.1
applied with the operators $T_k$.

\bigskip

\noindent {\it Notations.}  $\Omega_{ \{ \lambda \} }$ denotes
the ground state of the space  ${\cal H}_{ \{ \lambda \} }$
associated to the corresponding creation and annihilation operators
$a_{\lambda}$ and $a_{\lambda}^{\star}$.  One
knows that ${\cal H}_{ \{ \lambda \} }$ is associated with the orthonormal
basis $(h_j)_{(j\geq 0)}$ defined by:
 %----
 $$ h_0 = \Omega _{ \{ \lambda \} } \hskip 1cm h_{j+1} =  (j+1 )^{-1/2}\
a_{\lambda}^{\star} h_j$$
%-----
When identifying ${\cal H}_{ \{ \lambda \} }$ with $L^2(\R)$,
this basis is the basis of Hermite's functions and
$a_{\lambda}h_j = \sqrt {j} h_{j-1}$ ($j\geq 1$). We shall use the
following notations for the  operators belonging to the tensorial product
${\cal H }_F = {\cal H}_E \otimes {\cal H}_{ \{ \lambda
\} }$. We set $A = I \otimes a_{\lambda}$, $A^{\star} = I
\otimes a_{\lambda}^{\star}$ and for all $T\in {\cal L}({\cal
H}_F)$ we set $R(T) = \rho_{FE} (T) \otimes I $ where $ \rho
(T)$ is defined in section 2 by $ \rho (T) =
\pi_{EF}^{\star } T \pi_{EF}$.  Thus  $R(T) =i_{EF}  \rho_{FE}
(T)$.  In order to generalize the operator $\pi_{EF}$ we define for all
$j\geq 0$ an $\Psi _j$ from
${\cal H}_{E}$ into ${\cal H}_F$ by
%----
$$(\Psi _j f) = f \otimes h_j  $$
%----
We denote by
$\Psi_j^{\star}$ the adjoint operator of ${\cal H}_{F}$ in
${\cal H}_{E}$.
\smallskip
With these notations, we can sum up some of the usual properties on
Hermite's functions with the next lemma:

\bigskip
\noindent {\bf Lemma A.2.} {\it With these notations one has:
%----
$$ \sum _{j= 0}^{\infty} \Psi_j \Psi_j^{\star} = I \hskip 1cm
 \sum _{j= 0}^{\infty} \Vert\Psi_j^{\star}f \Vert _{{\cal H}_E}^2 =
 \Vert f \Vert _{{\cal H}_F}^2  \hskip 1cm \forall f\in {\cal H}_F
 \leqno (A.2)$$
 %------
 If we denote by ${\cal H}^m (E , F)$ ($m\geq 0$) the partial Sobolev
space consisting of the
 $f\in {\cal H}_F$ such that:
 %---
 $$ \Vert f \Vert _{{\cal H}^m (E , F)}^2 =  \sum _{j= 0}^{\infty}
  (1+j)^m \Vert\Psi_j^{\star}f \Vert _{{\cal H}_E}^2 < \infty,$$
 %---
then the operator $A A^{\star}$ with the domain ${\cal H}^2 (E , F)$ is
self-adjoint and verifies $A A^{\star} \geq I$. One has for all
$\alpha \in \R$ and for all $j\geq 0$:
%----
$$ (A A^{\star} )^{\alpha} \Psi_j = (j+1)^{\alpha } \Psi_j \hskip
1cm \Psi_j^{\star}  (A A^{\star} )^{\alpha}  = (j+1)^{\alpha }
\Psi_j^{\star} \leqno (A.3)$$
%------
We have for all $j\geq 1$:
%----
$$ A \Psi_j = \sqrt {j} \Psi_{j-1} \hskip 1cm \Psi_j^{\star }
A^{\star} = \sqrt {j} \Psi_{j-1}^{\star} \leqno (A.4)$$
%-----
(if $j=0$ then the right hand-sides are replaced by $0$.)  For all $j\geq
0$, we have:
%------
$$ A^{\star} \Psi_j =  \sqrt {j+1} \Psi _{j+1}
 \hskip 1cm
 \Psi_j^{\star} A = \sqrt
 {j+1} \Psi _{j+1}^{\star} \leqno (A.5) $$
 %----
}

\bigskip

For each operator $T$
 in ${\cal L}({\cal H}_{F})$ we define an operators-valued matrix
 $a_{jk}(T)$  in ${\cal L}({\cal H}_{E})$ by:
 %----
$$a_{jk}(T) = \Psi_j^{\star} T \Psi _k \leqno (A.6) $$
%-----
Thus $\pi_{EF}= \Psi _0$  and $\rho_{FE} (T) = A_{00}(T)$. The
norm of an operator $T$ in ${\cal L}({\cal H}_{F})$ may be estimated
starting from those of the $a_{jk}(T)$ using the following proposition
which is a  variant of Schur's Lemma.
  \bigskip

\bigskip
\noindent {\bf Proposition A.3.} {\it Set $T$ an element of
${\cal L}({\cal H}_{F})$. Suppose that there exists $M>0$ such that,
for all $k\geq 0$ and for all $\varphi$ in ${\cal H}_E$:
%------
$$   \sum _{j\geq 0} \Vert a_{jk}(T) \varphi \Vert_{{\cal H}_{E}}
\leq M \Vert \varphi \Vert_{{\cal H}_{E}} \hskip 1cm
 \sum _{j\geq 0} \Vert a_{jk}(T^{\star} ) \varphi
\Vert_{{\cal H}_{E}} \leq M \Vert \varphi \Vert_{{\cal H}_{E}}
\leqno (A.7)$$
%-----
Then $\Vert T \Vert_{{\cal L}({\cal H}_{F})} \leq M$. }

\bigskip
\noindent {\it Proof.} From lemma A.2, for all
$f$ and $g$ in ${\cal H} _F$ one gets:
%----
$$ \big < Tf , g\big > = \sum _{jk} \big < a_{jk}(T) \Psi
_k^{\star} f \ ,\ \Psi_j^{\star}g \big >$$
%----
One has:
%-----
$$ \left |\big < a_{jk}(T) \Psi _k^{\star} f \ ,\ \Psi_j^{\star}g
\big > \right | \leq \Vert a_{jk}(T) \Psi _k^{\star} f \Vert \
\Vert \Psi_j^{\star}g \Vert $$
%-----
This scalar product may be bounded by:
%----
$$ \left |\big < a_{jk}(T) \Psi _k^{\star} f \ ,\ \Psi_j^{\star}g
\big > \right | \leq \Vert
 \Psi _k^{\star} f \Vert \ \Vert  a_{jk}(T) ^{\star}
 \Psi_j^{\star}g \Vert $$
 %----
Consequently:
 %----
$$ \left |\big < a_{jk}(T) \Psi _k^{\star} f \ ,\ \Psi_j^{\star}g
\big > \right | \leq \Big (  \Vert a_{jk}(T) \Psi _k^{\star} f
\Vert \ \Vert  \Psi _k^{\star} f \Vert \Big ) ^{1/2} \ \ \ \Big (
\Vert a_{jk}(T) ^{\star}  \Psi_j^{\star}g
 \Vert \ \Vert    \Psi_j^{\star}g \Vert \Big ) ^{1/2} $$
 %---
From Cauchy-Schwarz:
 %----
 $$ \Big | \big < Tf , g \big > \Big |^2 \leq
 \left [ \sum _{jk}  \Vert a_{jk}(T) \Psi _k^{\star} f \Vert \
\Vert  \Psi _k^{\star} f \Vert \right ] \ \  \left [ \sum _{jk}
 \Vert a_{jk}(T) ^{\star}  \Psi_j^{\star}g
 \Vert \ \Vert    \Psi_j^{\star}g \Vert \right ]$$
 %---
Noticing that $\big ( a_{jk}(T) \big )^{\star} = a_{kj}
(T^{\star})$ we obtain:
 %---
 $$ \Big | \big < Tf , g \big > \Big |^2 \leq M^2
 \left [ \sum _{k\geq 0} \Vert  \Psi _k^{\star} f \Vert ^2 \right ] \ \
 \left [ \sum _{j\geq 0}  \Vert \Psi _j^{\star} g \Vert ^2 \right ]
 \ \leq \ M^2 \Vert f \Vert_{{\cal H}_{F}} ^2 \Vert g \Vert_{{\cal H}_{F}}
^2$$
 %---
The proof of proposition A.3 is completed.

\hfill \carre
%---------

\bigskip

\noindent {\bf Proposition A.4.} {\it Set $T$ an element in
${\cal W}_1(F)$. Assume that there is $M>0$ satisfying for all
$\varphi $ in ${\cal H}_E$:
%------
$$ \sup _{k\geq 0}  \sum _{j\geq 0} {\Vert a_{jk}(T) \varphi
\Vert_{{\cal H}_{E}} \over \sqrt {(j+1)(k+1)}} \leq M \Vert
\varphi \Vert _{{\cal H}_{E}} \leqno (A.8)$$
%-----
$$ \sup _{k\geq 0}  \sum _{j\geq 0} {\Vert a_{jk}(T^{\star})
\varphi \Vert_{{\cal H}_{E}} \over \sqrt {(j+1)(k+1)}} \leq M
\Vert \varphi \Vert_{{\cal H}_{E}} \leqno (A.9)$$
%-----
Then
%-----
$$ \Vert Tf \Vert \leq M \sqrt {2} \Vert ( A _{\lambda}^{\star})^2
f \Vert \ +\ \sqrt{ 2} \Vert [ A _{\lambda}^{\star} , T] \Vert\
  \Vert f\Vert \leqno (A.10)$$
%-----
for all $f$ in ${\cal H}^2 (E , F)$.
}

\bigskip
\noindent {\it Proof.} Set $S$ the operator $S= (A
A^{\star})^{-1/2} T (AA^{\star})^{-1/2} $. By lemma A.2
we have:
%----
$$a_{jk} (S) = { a_{jk}(T)  \over \sqrt {(j+1)(k+1)}}$$
%-----
Under the hypotheses of the Proposition the operator $S$ is then bounded
in ${\cal H}_F$ with a norm $\leq M$. From
lemma A.2 and for all $g$ in ${\cal H}_F$, we get
%----
$$ \Vert (AA^{\star})^{-1/2} A^{\star} g\Vert ^2 = \sum _{j\geq 1}
{j \over j+1} \Vert \Psi_{j-1}^{\star}g\Vert ^2 \geq {1 \over 2}
\sum _{j\geq 0} \Vert \Psi_j^{\star}g\Vert ^2= {1 \over 2} \Vert
g\Vert^2$$
%----
Consequently, for all $f$ in ${\cal H}^2 (E , F)$:
%---
$$ \Vert Tf\Vert \leq \sqrt {2} \Vert (AA^{\star})^{-1/2}
A^{\star} Tf\Vert \leq \sqrt {2} \Vert [  A^{\star} , T ] \Vert \
\Vert f\Vert + \sqrt {2}   \Vert (AA^{\star})^{-1/2} T  A^{\star}
f\Vert$$
%----
Indeed the operator $(AA^{\star})^{-1/2}$ has a norm $\leq
1$. We have:
%----
$$  \Vert (AA^{\star})^{-1/2} T  A^{\star} f\Vert \leq \Vert S
(AA^{\star})^{+1/2} A^{\star} f\Vert \leq M \Vert
(AA^{\star})^{+1/2} A^{\star} f\Vert = M \Vert (A^{\star})^2
f\Vert $$
%-----
Consequently, inequality (A.10) thus follows.

\hfill \carre

\bigskip

We shall apply proposition A.4 to the operator $T - R
(T)$ noticing that $R(T)$ commutes with $A$ and $A^{\star}$.
The operator $R (T)$ is chosen such that $a_{00} (T - R
(T))=0$. Using commutators, we shall estimate all the others elements
$a_{jk} (T - R (T))$. This is the purpose of the next proposition.

\bigskip
\noindent {\bf Proposition A.5.} {\it  Under the hypotheses of
Proposition A.1, for all $k\geq 0$ and for all $\varphi$ in
${\cal H}_E$ we have:
%---
$$S_k (T , \varphi ) := \sum _{j\geq 0 } {
 \Vert a_{jk } (T - R (T)) \varphi \Vert
 \over \sqrt {(j+1)(k+1)}}\leq C
 \Vert \varphi \Vert \  \sum _{ 1 \leq \alpha + \beta \leq 2 }
 \Vert (ad P_{\lambda})^{\alpha} (ad Q_{\lambda})^{\beta} T \Vert _{{\cal
L}({\cal
 H})}\
\leqno (A.11) $$
%----
and a similar expression holds when replacing $T$ by $T^{\star}$. }

%--------
\bigskip
\noindent {\it Estimations of $S_0 (T , \varphi)$.} We shall prove that:
%----
$$ S_0(T , \varphi) \ \leq \Vert [A , T ] \Vert \ \Vert \varphi
\Vert  \leqno (A.12) $$
%-------
From  lemma A.2 (point A.5),  one sees for all $j\geq 1$ that:
%-----
$$ \sqrt {j} a_{j0} (T - R (T)) = \Psi_{j-1}^{\star} [A , T]
\Psi_0$$
%------
Since $ a_{00} (T - R (T))=0$, it is deduced using
(A.2) that:
%----
$$S_0 (T , \varphi) \leq \sum _{j=1}^{\infty } {1 \over \sqrt
{j(j+1)}} \Vert  \Psi_{j-1}^{\star} [A , T] \Psi_0 \Vert \ \leq
\left [\sum _{j=1}^{\infty} {1 \over j(j+1)} \right ]^{1/2} \
\left [ \sum _{j=1}^{\infty} \Vert  \Psi_{j-1}^{\star} [A , T]
\Psi_0 \varphi \Vert ^2 \right ]^{1/2}  $$
%-----
$$ \leq \Vert [A , T] \Psi_0 \varphi \Vert \ \leq \Vert [A , T]
\Vert \ \Vert \varphi \Vert $$
%-----
 Inequality (A.12) is therefore true.

\bigskip

\noindent {\it Recursion  between the $S_k(T , \varphi)$.}
 If $k\geq 1$ we shall prove that:
  %------
  $$ S_k (T , \varphi) \leq {k \over k+1}  S_{k-1}  (T , \varphi)
  +  {C \Vert \varphi \Vert \over k+1}\Big [ \ \Vert [A , T] \Vert +
  \Vert [A^{\star} , T] \Vert + \  \Vert \ [ A^{\star} ,  [A^{\star} ,
  T]\  \Vert    \ \Big ]  \leqno (A.13) $$
  %------
To this end, we use the fact that, if $1\leq j\leq k$ the we have from
(A.4) (A.5):
%-----
$$ \sqrt {k} a_{jk}  (T - R (T))= \sqrt {j}  a_{j-1 ,k-1}  (T - R
(T))\ + \Psi_j^{\star} [T , A^{\star}] \Psi_{k-1}$$
%----
If $j=0$ then the first term above has to be replaced by  $0$. If $j>k$
then we use:
%-----
$$ \sqrt {j} a_{jk}  (T - R (T))= \sqrt {k}  a_{j-1 ,k-1}  (T - R
(T))\ + \Psi_{j-1} ^{\star} [A , T] \Psi_{k}$$
%----
Then we are able to write $S_k(T , \varphi) \leq S'_k(T ,
\varphi) + S''_k(T , \varphi) + S'''_k(T , \varphi)$ where:
%----
$$S'_k(T , \varphi) = \sum _{j= 1} ^{\infty} \inf \left ( \sqrt
{j\over k} , \sqrt {k\over j} \right ) \ { \Vert a_{j-1, k-1 } (T
- R(T)) \varphi \Vert  \over \sqrt {(j+1)(k+1)}} \leq  {k \over
k+1} S_{k-1}  (T , \varphi)$$
%------
$$ S''_k(T , \varphi) =  \sum _{j= k+1} ^{\infty} { \Vert
\Psi_{j-1}^{\star} [T , A ] \Psi_k \varphi \Vert \over \sqrt {
j(j+1)(k+1)}} $$
%-----
$$ S'''_k(T , \varphi) =  \sum _{j= 0}^k  { \Vert \Psi_{j}^{\star}
[T , A^{\star} ] \Psi_{k-1} \varphi \Vert \over \sqrt {
(j+1)k(k+1)}} $$
%-----
From (A.2) and since $ \Vert  \Psi_k \varphi \Vert = \Vert
\varphi \Vert $:
%---
 $$ S''_k(T , \varphi)  \leq {1 \over \sqrt {k+1}}
\left [ \sum _{j=k+1}^{\infty} {1 \over j(j+1)} \right ]^{1/2} \
\left [ \sum _{j \geq 1}  \Vert \Psi_{j-1}^{\star} [T , A ] \Psi_k
\varphi \Vert^2 \right ]^{1/2} \leq {1 \over k+1} \Vert [T , A]
\Vert \ \Vert \varphi \Vert $$
%----
If $k=1$ then we see that  $ S'''_1(T , \varphi) \leq \Vert [A^{\star
}, T] \Vert \  \Vert \varphi \Vert $. If $k\geq 2$ then the estimation of
$ S'''_k(T , \varphi)$ involves commutators with length 2.
We still have, if $j\leq k$:
%-----
$$ \sqrt {k-1}  \Psi_{j}^{\star} [T , A^{\star} ] \Psi_{k-1} =
\sqrt {j} \Psi_{j-1}^{\star} [T , A^{\star} ] \Psi_{k-2} +
 \Psi_{j}^{\star} [[T , A^{\star} ],A^{\star} ]  \Psi_{k-2}$$
 %------
Consequently, if $k\geq 2$:
 %----
 $$ S'''_k(T , \varphi) \leq \sum _{j=1}^k \  \sqrt {j\over k-1} \
  {\Vert \Psi_{j-1}^{\star} [T ,
 A^{\star} ] \Psi_{k-2} \varphi \Vert  \over \sqrt
 {(j+1)k(k+1)}}+ ...$$
 %----
 $$ ...  + \sum _{j=0}^k { \Vert  \Psi_{j}^{\star} [[T , A^{\star}
],A^{\star} ]
 \Psi_{k-2} \varphi \Vert \over \sqrt
 {(j+1)(k+1)k(k-1)}}
 $$
 %------
Using again Cauchy-Schwarz and  lemma A.2, we obtain if $k\geq
 2$:
 %-----
 $$  S'''_k(T , \varphi) \leq {1 \over \sqrt{ k(k-1)}} \Big [ \Vert
 [A^{\star} , T] \Vert + \Vert [ A^{\star} , [ A^{\star} , T] ] \Vert
 \Big ] \ \Vert \varphi \Vert $$
 %--------
 We then deduce the validity of (A.13). Inequality (A.11) follows by
iteration on (A.12) and (A.13). The proposition A.1 is a consequence of
  Propositions A.4 and A.5 and the proof of
 Proposition 2.5 is finished.

 \bigskip
%---------APPENDICE--------BBBBB-------

 \noindent {\bf Appendice B. Differential systems.}
 \bigskip

\bigskip
\noindent {\bf Proposition B.1.} {\it Suppose that we are given for all
$\lambda $ and $\mu$ in $\Lambda _n$ a continuous map $t
\rightarrow \Omega _{\lambda \mu} (t)$ from $\R$ into ${\cal L}(
{\cal L}({\cal H}))$. %---
Assume that there are  $\gamma >0$ and $S_{\gamma} >0$ such that,
for all $\lambda$ and $\nu$ in $\Lambda_n$, for all $t\in
\R$:
%-----
$$ \sum _{\mu \in \Lambda_n} \Vert \Omega _{\lambda \mu} (t) \Vert
_{ {\cal L}( {\cal L}({\cal H}))}  \ e^{-\gamma |\mu - \nu|} \leq
S_{\gamma} e^{-\gamma |\lambda - \nu|}\leqno (B.1)$$
%-----
Then,  for all $s\in \R$, there exists functions $t
\rightarrow A_{\lambda \mu}^{(0)}  (t, s)$ and  $t \rightarrow
A_{\lambda \mu}^{(1)}  (t, s)$ ($(\lambda , \mu) \in \Lambda_n^2$)
being $C^1$ from $\R$ into ${\cal L}( {\cal
L}({\cal H}))$ such that:
%-------
$$ {d\over dt} A_{\lambda \mu}^{(0)}  (t, s) =  A_{\lambda
\mu}^{(1)} (t , s) \hskip 1cm  {d\over dt} A_{\lambda \mu}^{(1)}
(t, s ) = \sum _{\nu \in \Lambda_n}  \Omega _{\lambda \mu} (t)
\circ A_{\nu \mu}^{(0)}  (t, s)\leqno (B.2)$$
%-------
$$   A_{\lambda  \mu} ^0 (s, s) = \delta _{\lambda \mu } I \hskip
1cm
 A_{\lambda  \mu} ^1 (s, s) = 0 \leqno (B.3)$$
 %------
(in (B.1), the composition is the one of ${\cal L}( {\cal L}({\cal
H}))$ and in (B.3) the identity operator $I$ is the one of ${\cal L}(
{\cal L}({\cal H}))$.) Moreover, if $M> \sqrt {S_{\gamma}}$, there exists
$C(M, \gamma) >0$ independent of $n$ such that:
%----
 $$ \Vert A_{\lambda \mu}^{(j)}  (t, s) \Vert_{ {\cal L}( {\cal L}({\cal
H}))} \leq C(M, \gamma)\ e^{M|t-s|} e^{-\gamma |\lambda- \mu|}
\hskip 1cm \forall (\lambda , \mu) \in \Lambda_n^2 \leqno (B.4)$$
%-----
There are also operator-valued matrices $t \rightarrow
B_{\lambda \mu}^{(0)}  (t, s)$ and  $t \rightarrow B_{\lambda
\mu}^{(1)}  (t, s)$ satisfying the same system together with the same
estimations and the same initial conditions:
%-----
$$ B_{\lambda  \mu} ^0 (s, s) = 0 \hskip 1cm
 B_{\lambda  \mu} ^1 (s , s) = \delta _{\lambda \mu } I \leqno (B.5)$$
 %------
}

\bigskip

  \noindent {\it Proof.} Let $E_{ n \gamma}$ be the set of all matrices
  $A= (A_{\lambda , \mu}) _{( (\lambda , \mu)\in \Lambda_n^2)}$ where each
  $A_{\lambda , \mu}$ is a map in ${\cal L}({\cal
  L}({\cal H}))$ which is  associated with the norm:
  %----
  $$ \Vert A \Vert _{n , \gamma} = \sup _{ (\lambda , \mu) \in
  \Lambda _n^2} e^{ \gamma |\lambda - \mu |} \ \Vert A_{\lambda \mu}
  \Vert _{ {\cal L}({\cal  L}({\cal H}))} $$
  %-----
  The left composition by the operators-valued matrix
  $\Omega _{\lambda \mu} (t)$ defines a map $\Omega (t)$
  in ${\cal L} ( E_{ n \gamma} )$ with a norm $\leq S_{\gamma}$.  For all
$\varepsilon >0$ we can associate to $E_{ n \gamma} ^2$ a  norm
  such that
  %---
  $$ U(t) = \pmatrix { 0 & I \cr \Omega (t) & 0 \cr } $$
  %-----
  is $\leq \sqrt { S_{\gamma} } (1 + \varepsilon)$. The stated result is
then valid.

  \bigskip

  \noindent {\it Remark 1.}  In the tensorial product
  $(E_{ n \gamma}^2) \otimes (E_{ n \gamma}^2) $ let
  $V(t)$ be the map defined by $V(t)  = U(t)  \otimes I  \ +
  \ I \otimes U(t)$. For all $\varepsilon>0$ one may
 associate  $(E_{ n \gamma}^2) \otimes (E_{ n \gamma}^2) $ with a norm
such that the map $V(t)$ is $\leq
 2 \sqrt { S_{\gamma} } (1 + \varepsilon)$. Consequently, if
 $M> 2  \sqrt { S_{\gamma} }$ and if $A_0$ is in
 $(E_{ n \gamma}^2) \otimes (E_{ n \gamma}^2) $ then the differential
system:
 %-----
 $$ A'(t) = U(t) A(t) \hskip 1cm   A(0) = A_0 $$
 %----
 has a solution taking values into  $(E_{ n \gamma}^2) \otimes (E_{ n
 \gamma}^2) $ and with an time exponential growth  $e^{M|t |}$.

 \bigskip

 \noindent
 {\it Remark 2.} If we are also given the continuous functions
$t\rightarrow F_{\lambda
 } (t)$ from
  $\R$ and taking values into ${\cal L}({\cal H})$ then the
 family of functions $t\rightarrow X_{\lambda}^{(j)} (t)$
 defined by:
 %-----
 $$ X_{\lambda}^{(j)} (t) =\sum _{\mu \in \Lambda_n}  \int _0^t
 B_{\lambda  \mu} ^j (t, s) \Big (  F_{\mu } (s)\Big ) ds
 $$
%------
satisfies the differential system:
%------
$$ {d\over dt} X_{\lambda}^{(0)} (t) = X_{\lambda}^{(1)} (t)
\hskip 1cm {d\over dt} X_{\lambda}^{(1)} (t) = \sum _{\mu \in
\Lambda_n} \Omega _{\lambda \mu} (t) \Big ( X_{\mu}^{(0)} (t)\Big
)\ + \ F_{\lambda}(t)$$
%----
together with the initial conditions $X_{\lambda}^{(j)} (0) = 0$ and the
following estimates (for example if $t>0$):
%------
$$ \Vert X_{\lambda}^{(j)} (t)\Vert _{{\cal L}({\cal H})} \leq C
(M, \gamma) \sum _{\mu \in \Lambda_n} e^{-\gamma |\lambda- \mu|}
\int_0^t e^{M|t-s|} \Vert  F_{\mu } (s)\Vert_{{\cal L}({\cal H})}
\ ds$$
%-----

\vfill \eject

%XXXXXXXXXXXXXXXXXXXXXXXXXXXXXXXXXX
%XXXXXXXXXXXXXX-BIBLIO-XXXXXXXXXXXXX

\noindent {\bf References.}
 \bigskip
 %-----
 \noindent [A-C-L-N] L. Amour, C. Cancelier, P. L\'evy-Bruhl, J.
 Nourrigat, {\it Decay of quantum correlations on a lattice by
 heat kernel methods}, Ann. Henri Poincar\'e, {\bf 8}, (2007),
 1469-1506.
 \medskip
 %--------
\noindent [A-J-P] Open quantum systems. I. The Hamiltonian
approach. Lecture notes from the Summer School held in Grenoble,
June 16--July 4, 2003. Edited by S. Attal, A. Joye and C.-A.
Pillet. Lecture Notes in Mathematics, 1880. Springer-Verlag,
Berlin, 2006.
\medskip
%---------
 \noindent [BE] R. Beals, {\it Characterization of pseudo-differential
 operators and applications.}, Duke Math. Journal, {\bf 44}, (1977),
 p.45-57.
 \medskip

%--------
 \noindent [BO1] J.M. Bony, {\it Evolution equations and microlocal
 analysis,} in {\it Hyperbolic Problems and related topics},
 (Conference in Cortona, 2002, Colombini and Nishitani, ed) , p.
 17-40.   Graduate series in Analysis, Int. Press, 2003.
\medskip
%------
\noindent [BO2] J.M. Bony, {\it Analyse microlocale et \'equations
d'\'evolution.}  S\'eminaire Equations aux d\'eriv\'ees
partielles, 2006-2007, Exp. XX, 14p. Ecole Polytechnique,
Palaiseau 2007.
\medskip
%------
\noindent [B-R] O. Bratteli, D.W. Robinson, {\it Operators
Algebras and Quantum Statistical Mechanics}, vol I and II,
Springer.
\medskip
%--------
\noindent [C-T] Cohen-Tannoudji, {\it M\'ecanique quantique,}, t.
I and II, Hermann (Paris), 1997.
\medskip
%----------
\noindent [C-V] A.P. Calderon, R. Vaillancourt, On the boundedness
of pseudodifferential operators, {\it J. Math. Soc. Japan}, {\bf
23}, 374-378, (1971).
\medskip
%-------
\noindent [D-J-P]  J. Derezi\'nski, V. Jaksi\'c, C.-A. Pillet,
{\it Perturbation theory of $W^{\star}-$dynamics, Liouvilleans and
KMS-states.}
\medskip
%--------
\noindent [HE] B. Helffer, {\it Th\'eorie spectrale pour des
op\'erateurs globalement elliptiques.} Ast\'erisque 112, S.M.F.
Paris, 1984.
\medskip
%--------
\noindent [HO] L. H\"ormander, {\it The analysis of linear Partial
differential operators.} Springer-Verlag, Berlin, 1990.
\medskip
%----------------
\noindent [L1] B. Lascar, Noyaux d'une classe d'op\'erateurs
pseudo-diff\'erentiels sur l'espace de Fock. {\it S\'eminaire Paul
Kr\'ee, Equations aux d\'eriv\'ees partielles en dimension
infinie,} Expos\'e 6, 1976-1977.
\medskip
%----------------
\noindent [L2] B. Lascar, Equations aux d\'eriv\'ees partielles en
dimension infinie, {\it Vector spaces measures and applications,
Proc. Conf. Univ. Dublin, 1977}, vol. I, Springer, 1978.
\medskip
%----------
\noindent [LI-R] E. Lieb, D.W. Robinson, {\it The finite group
velocity of quantum spin systems,} Comm. Math. Phys. {\bf 28}
(1972), 251-257.
\medskip
%----------
\noindent [MA-MI] V.A. Malyshev, R.A. Minlos, {\it Linear infinite
particle operators,} Translations of Mathematical Monographs, {\bf
143}, A.M.S, Providence (R.I.), 1995.
\medskip
%------
\noindent [MI-V-Z] R.A. Minlos, A. Verbeure, V.A. Zagrebnov, {\it
A quantum crystal model in the light-mass limit: Gibbs states},
Rev. Math. Physics, {\bf 12}, (7), (2000), 981-1032.
\medskip
%-------------
 \noindent [N-S]  B. Nachtergaele, R. Sims, {\it
Lieb-Robinson bounds and the exponential clustering theorem.}
Comm. Math. Phys. {\bf 265}, 119-130, (2006).
\medskip
%------
\noindent [N-O-S]  B. Nachtergaele, Y. Ogata, R. Sims, {\it
Propagation of correlations in quantum larttice systems,} J. of
Stat. Physics, {\bf 124}, 1, july 2006.
\medskip
%------
\noindent [N-R-S-S]  B. Nachtergaele,  H. Raz, B. Schlein, R.
Sims, {\it Lieb-Robinson bounds for harmonic and anharmonic
lattice systems,} Comm.in Math. Phys, {\bf 286} (3) (2008),
1073-1098. DOI 10.1007/s00220-008-0630-2.
\medskip
%------
\noindent [RO] D. Robert, {\it Autour de l'approximation
semi-classique.} Progress in Mathematics, 68, Birkh\"auser,
Boston, 1987.
\medskip
%---------
\noindent [Re-SI] M. Reed, B. Simon, {\it Methods of modern
Mathematical Physics,} vol I and II, Academic Press, 1975.
\medskip
%--------
\noindent [SH] M.A. Shubin, {\it Pseudodifferential operators and
spectral theory.} Springer-Verlag, Berlin, 1980.
\medskip
%------
\medskip
\noindent [SI] B. Simon,  {\it The statistical Mechanics of
lattice gases. } Vol. I. Princeton Series in Physics. Princeton,
1993.
%--------
\medskip
\noindent [TH] W. Thirring, {\it Quantum Mathematical Physics,
Atoms, Molecules and Large Systems,}  Second edition, Springer,
2002.
\medskip
%------

%--------------THE------END------------
\end